\documentclass[a4paper,11pt]{article}

\usepackage[utf8]{inputenc}
\usepackage{amsfonts}
\usepackage{dsfont}
\usepackage{tcolorbox}
\usepackage{xparse}

\usepackage{amsmath}
\usepackage{amssymb}
\usepackage{mathtools}
\usepackage{color}
\usepackage{bbold}

\makeatletter
\newsavebox{\@brx}
\newcommand{\llangle}[1][]{\savebox{\@brx}{\(\m@th{#1\langle}\)}%
  \mathopen{\copy\@brx\kern-0.5\wd\@brx\usebox{\@brx}}}
\newcommand{\rrangle}[1][]{\savebox{\@brx}{\(\m@th{#1\rangle}\)}%
  \mathclose{\copy\@brx\kern-0.5\wd\@brx\usebox{\@brx}}}
\makeatother

\pdfoutput=1 

\usepackage{jheppub} 

\usepackage[T1]{fontenc}

\title{\boldmath Symplectic Grassmannians, dual conformal symmetry and 4-point amplitudes in 6D}

\author[]{Klaus Bering,}
\author[]{Michal Pazderka}

\affiliation[]{Institute of Theoretical Physics \& Astrophysics, Masaryk University, Kotlarska 2, CZ-611 37 Brno, Czech Republic}

\emailAdd{bering@physics.muni.cz}
\emailAdd{michpa@mail.muni.cz}

\abstract{We investigate a new algebra-based approach of finding Grassmannian formulas for scattering amplitudes. Our prime motivation  
is massive amplitudes of 4D $\mathcal{N}=4$ SYM, and therefore we consider a 6D Grassmannian formula, where we can take advantage of massless kinematics. We next use symmetry arguments, and in particular, 6D dual conformal symmetry generalized to arbitrary dual conformal weights.
Assuming a rational ansatz in terms of Pl\"{u}cker coordinates (i.e.\ minors) for the integrand, this approach leads to a set of algebraic equations.
As an example, we explicitly 
find the solution for 4-point scattering amplitudes up to proportionality constants.}


\begin{document} 
\maketitle
\flushbottom

\section{Introduction}

Scattering amplitudes in various quantum field theories (QFTs) are important from both a theoretical and an experimental point of view.
It is, however, extremely challenging to calculate higher point amplitudes using Feynman diagrams because of a rapidly increasing number of diagrams for more external legs and/or more loops. Although the number of diagrams grows rapidly, a huge cancellation between terms might appear and the final result is simple, cf. e.g.\ \cite{Parke_Taylor}. Thus new structures among such amplitudes are expected, as well as new methods of calculating scattering amplitudes.

One recent such big discovery, the amplituhedron, relates scattering amplitudes of $\mathcal{N}=4$ SYM and positive geometry  \cite{Hamed_Trnka13}. Subsequently, many of such positive geometries has been discovered, e.g. \cite{Hamed17-associahedron, Hamed17-Cosmology}. A crucial part of the original discovery was a rewriting of scattering amplitudes as a sum over residues of certain integrals over Grassmannians. This Grassmannian representation of scattering amplitudes has been found for many theories, e.g. 4D $\mathcal{N}=4$ SYM, 3D ABJM theory, supergravity (SUGRA), etc.\ \cite{Hamed09, Lee10, Lipstein20,Heslop16}. However all such descriptions are naturally connected to massless kinematics.

We focused in this paper on a generalisation of the Grassmannian representation of scattering amplitudes for massive particles in 4D using symplectic Grassmannians. Although many of the discussed features will be general, our prime motivation is
 scattering amplitudes of 4D massive $\mathcal{N}=4$ SYM on the Coulomb branch \cite{Elvang11}. This theory is widely studied in the literature and some symplectic Grassmannian formulas based on rational maps can be found in literature \cite{Cachazo, Schwarz}. However, there the integrands are only expressed in terms of world-sheet coordinates while the integral representation in terms of Pl\"{u}cker coordinates is missing. 

A central role will be played by the symmetries of the theory. Unlike massless $\mathcal{N}=4$ SYM, the massive $\mathcal{N}=4$ SYM on the Coulomb branch is not a conformal field theory, but it still enjoys dual conformal symmetry \cite{Plefka14}. The (super)conformal symmetry breaking has deep consequences on the form of the Grassmannian representation. "Pure" Grassmannian formulas, such as those of massless 4D $\mathcal{N}=4$ SYM and massless 3D ABJM theory, naturally talks to the (super)conformal structure. Thus a new ansatz for QFTs without conformal symmetry but with dual conformal symmetry will be proposed and the implications of dual conformal generator will be investigated.

The paper is organized as follows. Chapter 2 reviews spinor-helicity kinematics in 3D, 4D \& 6D to fix notation and conventions. The advantages and disadvantages of using  the chiral model are discussed in chapter 3. Because of similarities between the known Grassmannians, which are summarized in chapter 4, we propose a 6D Grassmannian formula based on symplectic Grassmannian and deduce some of its properties via symmetries of massive 4D $\mathcal{N}=4$ SYM.  Due to its importance, the implications of special dual conformal generator are investigated in chapter 6. The results obtained in all previous chapters are applied to the 4-point example in chapter 7, which is immediately followed by comparing with known 6D results in chapter 8.

\section{Kinematical data and formalism}
\label{kinematical data}
Before reviewing the Grassmannian formulas it is useful to discuss the kinematical data and their notation. By "kinematical data" is meant a collection of (i) Grassmann-even spinors describing external particle momenta and (ii) Grassmann-odd spinors that parametrize the supersymmetry.

This paper contains examples in three various spacetime dimensions: 3D, 4D and 6D. Bosonic spinor variables can be conveniently grouped into 2- or 4-planes, which goes hand in hand with the interpretation of Grassmannians as $k$-planes. Let us now briefly recapitulate this connection.
\begin{itemize}
    \item \textbf{3D}: A massless vector in 3D can be decomposed as a product of two spinors $\lambda_i^{\alpha}$
    \begin{equation}
        p_i^{\alpha\beta}~=~\lambda^{\alpha}_i\lambda^{\beta}_i~,
    \end{equation}
    where the index $\alpha=1,2$ denotes a spinor representation of the 3D Lorentz group $Spin(2,1)\cong SL(2,\mathbb{R})$ and the index $i=1,\ldots, n$ is a particle index. Such spinors can now be arranged into a $2\times n$ matrix 
    \begin{equation}
        \Lambda^{\alpha}_{i}~:=\left(
        \begin{array}{cccc}
             \lambda^{\alpha}_1&\lambda^{\alpha}_2&\ldots&\lambda^{\alpha}_n  
        \end{array}\right)_{2\times n}~,
    \end{equation}
    which can be viewed as a 2-plane in an $n$-dimensional space. It is an element of the orthogonal Grasmannian $OG(2,n)$ because of momentum conservation.
    
    \item \textbf{4D}: Similarly a massless vector in 4D can be decomposed into a product of two  distinct spinors $\lambda^{\alpha}_i$ and $\tilde{\lambda}^{\dot{\alpha}}_i$ 
    \begin{equation}
        p^{\alpha\dot{\alpha}}_i~=~\lambda^{\alpha}_i\tilde{\lambda}^{\dot{\alpha}}_i~; 
    \end{equation}
    where indices $\alpha=1,2$ and $\dot{\alpha}=1,2$ are in $SL(2,\mathbb{C})\times SL(2,\mathbb{C})\cong Spin(4,\mathbb{C})$, that is, the complexified spin group of the 4D Lorentz group $SO^{+}(1,3)$; and $i=1,\ldots ,n$ is the particle index. These spinors can now be also compactly written as two $2\times n$ matrices
    \begin{equation}
        \Lambda^{\alpha}_i~:=\left(
        \begin{array}{cccc}
             \lambda^{\alpha}_1&\lambda^{\alpha}_2&\ldots&\lambda^{\alpha}_n  
        \end{array}\right)_{2\times n}~, \qquad
        \tilde{\Lambda}^{\dot{\alpha}}_i~:=\left(
        \begin{array}{cccc}
            \tilde{\lambda}^{\dot{\alpha}}_1&\tilde{\lambda}^{\dot{\alpha}}_2&\ldots&\tilde{\lambda}^{\dot{\alpha}}_n  
        \end{array}\right)_{2\times n}~,
    \end{equation}
    which can be viewed as elements of two Grassmanians orthogonal to each other. 
    
    \item \textbf{6D}: A 6D massless vector can be decomposed also as a product of spinors $\lambda^{Aa}_i$ but now with three indices 
    \begin{equation}
        p^{AB}_i~=~\lambda^{Aa}_i\lambda^{B}_{ia}~=~\lambda^{Aa}_i\epsilon_{ab}\lambda^{Bb}_i~;
    \end{equation}
    where spinor indices $A,B=1,2,3,4$ are in $SL(4,\mathbb{C})\cong SU(4)_{\,\mathbb{C}}\cong Spin(6,\mathbb{C})$, that is, the complexified spin group of the 6D Lorentz group $SO^{+}(1,5)$; $a,b=1,2$ are indices of one chiral $SU(2)$ of the little group $[SU(2)\times SU(2)]/\mathbb{Z}_2\cong SO(4)$; and $i=1,\ldots ,n$ is the particle index. We use the convention that $\epsilon_{12}=1$. Spinors can be compactly written as a matrix 
    \begin{equation}
        \Lambda^{A\mu}~:=\left(
        \begin{array}{cccccc}
             \lambda^{A1}_1&\ldots&\lambda^{A1}_n&\lambda^{A2}_1&\ldots&\lambda^{A2}_n  
        \end{array}
    \right)_{4\times 2n}~,
    \end{equation}
    \begin{equation}
    \Lambda^{A}{}_{\mu}:=\lambda^{A\nu}\Omega_{\nu\mu}~, \qquad \Omega_{\mu\nu}=\left(
    \begin{array}{cc}
         0&\mathbb{1}_{n\times n}  \\
         -\mathbb{1}_{n\times n}&0 
    \end{array}
    \right)_{2n\times 2n}~,
    \end{equation}
    where the index $\mu=(i,a)$ is a double index containing particle index $i$ and little group index $a$.
(Unfortunately, the double index notation $\mu=(i,a)$ becomes tedious 
whenever incomplete sums of the type $\sum_{i<j}$ over particle indices appear in the following. In contrast, the little group index is always fully summed over. Thus, we shall use this double index notation when convenient, but not always.)  The $4\times 2n$ matrix $\Lambda^{A\mu}$ can be viewed as an element of Grassmannian $Gr(4,2n)$. The conservation of momenta now implies that the Grassmanian is a symplectic Grassmannian ${LG}(4,2n)$.
\end{itemize}

In case of superamplitudes the external data contains not only the Grassmann-even Lorentz spinors $\lambda_i$ but Grassmann-odd parameters $\eta_i$ as well. Therefore it is convenient to use a condensed superization notation \cite{6D_nase}
\begin{equation}
\label{superize}
    \mathcal{A}=(A,I) \quad \Rightarrow\quad \Lambda^{\mathcal{A}}_{ia}=(\Lambda^{A}_{ia},\eta^I_{ia} )~.
\end{equation}

\section{Chiral toy model}

We consider just the chiral model of full 6D space in this work. In other words, we use $\lambda^{Aa}_i$ only instead of a pair $\lambda^{Aa}_i$ and $\tilde{\lambda}_{iA\dot{a}}$ related by 
\begin{equation}
    \lambda^{Aa}_i\lambda^{B}_{ia}=\frac{1}{2}\epsilon^{ABCD}\tilde{\lambda}_{iC\dot{a}}\tilde{\lambda}_{iD}^{\dot{a}}~.
\end{equation}
The reader may wonder why we ignore half of the 6D kinematical variables. The reason for this is in our focus on massive 4D amplitudes. Let us discuss how this works. To get from 6D theory to 4D theory we use dimensional reduction, where the used embedding of 4D spinors into 6D spinors is \cite{Plefka14}
\begin{equation}
\label{dim_red_4D}
    \lambda^{Aa}_i=\left(
    \begin{array}{cc}
         -\mu_{\alpha}& \lambda_{\alpha} \\
         \tilde{\lambda}^{\dot{\alpha}}&\tilde{\mu}^{\dot{\alpha}} 
    \end{array}\right)~, \qquad \tilde{\lambda}_{A\dot{a}}=\left(
    \begin{array}{cc}
         \bar{\rho}\mu^{\alpha}&\lambda^{\alpha}  \\
         -\tilde{\lambda}_{\dot{\alpha}}& \rho\tilde{\mu}_{\dot{\alpha}}  
    \end{array}
    \right)
    \quad \textrm{with}\quad \rho=\bar{\rho}^{-1}=\frac{m}{\bar{m}}~.
\end{equation}
Here spinors are related to massive 4-momenta as $p^{\alpha\dot{\alpha}}=\lambda^{\alpha}\tilde{\lambda}^{\dot{\alpha}}+\mu^{\alpha}\tilde{\mu}^{\dot{\alpha}}$ and satisfy $\langle\lambda\mu\rangle:=\lambda^{\alpha}\epsilon_{\alpha\beta}\mu^{\beta}=m$, $[\tilde{\mu}\tilde{\lambda}]:=\tilde{\mu}_{\dot{\alpha}}\epsilon^{\dot{\alpha}\dot{\beta}}\tilde{\lambda}_{\dot{\beta}}=\bar{m}$ and $p^2=m\bar{m}$. If we restrict ourselves to the case $m=\bar{m}$, the main difference between the dimensionally reduced spinors $\lambda^{Aa}$ and $\tilde{\lambda}_{A\dot{a}}$ in eq.\ \eqref{dim_red_4D} is in the position of 4D spinor indices.  This reflects the fact that we cannot raise or lower individual spinor indices in 6D! This is, however, possible in 4D, because in 4D there exist Levi-Civita tensors $\epsilon_{\alpha\beta}$ and $\epsilon_{\dot{\alpha}\dot{\beta}}$ to raise or lower individual spinor 
indices. This can be shown on the Grassmann $\delta$-function present in the amplitudes
\begin{equation}
    \delta^{4}(q^{I\alpha})=\prod_{\alpha=1}^2\prod_{I=1}^2q^{I\alpha}=\left[\prod_{I=1}^2q^{I1}\right]\left[\prod_{I=2}^2q^{I2}\right]=\left[\prod_{I=1}^2(-q^{I}_2)\right]\left[\prod_{I=1}^2q^{I}_1\right]=\prod_{\alpha=1}^2\prod_{I=1}^2q^I_{\alpha}=\delta^{4}(q^{I}_{\alpha})~.
\end{equation}
Therefore, from a 4D perspective, it is sufficient to consider $\lambda^{Aa}_i$ only. 
There is, however, one issue. We are no longer able to construct two-spinor Lorentz invariants $\langle i^{a}|j_{\dot{b}}]:=\lambda^{Aa}_i\tilde{\lambda}_{jA\dot{b}}$, but only 
\begin{equation}
\label{6D_anglebracket}
    \langle i j k l \rangle:=\epsilon_{ABCD}\lambda^{Aa}_i\lambda^{B}_{ja}\lambda^{Cc}_k\lambda^D_{lc}~.
\end{equation}
Consequently, we will only be able to construct 6D amplitudes of even particle number.

There is an analogous question on the relation between the 4D little group $SU(2)$ and the 6D little group $SU(2)\times SU(2)$. It was shown in \cite{Plefka14, Schwarz} that only one of the $SU(2)$ factors survives the dimensional reduction \eqref{dim_red_4D} of the 6D helicity generators
\begin{equation}
    h_{ab}=\sum_{i=1}^n[\lambda^{A}_{i(a}\partial_{iAb)}-\xi_{i(a}\partial_{ib)}]~, \qquad \tilde{h}_{\dot{a}\dot{b}}=\sum_{i=1}^n[\tilde{\lambda}_{iA(\dot{a}}\partial^A_{i\dot{b})}-\tilde{\xi}_{i(\dot{a}}\partial_{i\dot{b})}]~.
\end{equation}
It turns out that under the dimensional reduction \eqref{dim_red_4D} the $SU(2)$ factors become identified. This is because the 4D massive little group is diagonally embedded $SU(2)\ni g ~\hookrightarrow~ (g,g) \in SU(2)\times SU(2)$ into the 6D little group. As a consequence, a single $SU(2)$ is sufficient in the 6D chiral toy model to describe 4D massive kinematics.

Lastly, let us discuss the Grassmann parametrisation. Scattering amplitudes of massive 4D $\mathcal{N}=4$ SYM on the Coulomb branch are conventionally expressed in non-chiral (2,2) superspace. The main idea is to Fourier-transform half of the Grassmann parameters $\eta^{I}$ of the $\mathcal{N}=4$ on-shell superspace. 
(Be aware that the literature differs on which components of $\eta^{I}$ are transformed.) We use the convention of \cite{Plefka14} mainly, where instead of  $\{\eta^2,\eta^3\}$ we use the non-chiral Fourier-transform $\{\tilde{\eta}_2,\tilde{\eta}_3\}$. These are conventionally packaged into two objects, e.g.
\begin{equation}
    \xi_a:=(\tilde{\eta}_{3},\eta^1)~, \qquad \tilde{\xi}^{\dot{a}}:=(\tilde{\eta}_2, -\eta^{4})~. 
\end{equation}
We use the "chiral" version, where non-chiral Grassmann parameters are grouped as
\begin{equation}
\label{eta_embeding}
    \eta^{Ia}_{i}:=\left(
    \begin{array}{cc}
        \eta^1_i & -\tilde{\eta}_{i3} \\
         -\eta^{4}_i&\tilde{\eta}_2 
    \end{array}
    \right)~.
\end{equation}
Similar packaging of $\xi_{a}$ and $\tilde{\xi}^{\dot{a}}$ has been also used in \cite{Schwarz}.

\section{Review on Grassmannian formulas}
\label{review}

Let us briefly discuss the known Grassmannian formulas for scattering amplitudes, Grassmannian geometries and their relation to kinematics in appropriate dimensions. 

\subsection{3D and 4D Grassmannian formulas}
The 3D example is ABJM theory \cite{Lee10, Huang13}. The Grassmannian representation of scattering amplitudes is based on a so-called orthogonal Grassmannian $C\in {OG}(k,2k)$. The orthogonal Grassmannian can be viewed as a $k\times 2k$ matrix $C_{m i}$ satisfying 
\begin{equation}
    (C C^T)_{mp}=\sum_{i=1}^{2k}C_{m i}C_{p i}=0~.
\end{equation}
A tree-level scattering amplitudes of ABJM theory can now be written as 
 \begin{equation}
 \label{3D_grass}
        \mathcal{L}_{k,2k}=\int \frac{d^{k\times 2k}C}{\textrm{Vol}(GL(k))}\frac{1}{M_jM_{j+1}\ldots M_{j+k-1}}\delta^{\frac{k(k+1)}{2}}(CC^T)\prod_{m=1}^k\delta^{2|3}(C_m \Lambda^T)~,
    \end{equation}
where $M_j$ is a minor composed of $k$ consecutive columns from $j$ to $j+k$. 

The ABJM theory is a conformal field theory \cite{Bargheer10, Lipstein11}. Thus there is a dilaton generator of the form
\begin{equation}
    d_{3D}=\sum_i\left[\frac{1}{2}\lambda^{\alpha}_i\frac{\partial}{\partial \lambda^{\alpha}_i}+\frac{1}{2}\right]~,
\end{equation}
which annihilates \eqref{3D_grass}. This can be easily shown by acting with the Euler operator on the bosonic part of the $\delta$-function.

The 4D example is massless $\mathcal{N}=4$ SYM theory. The Grassmannian  formula for planar tree-level scattering amplitudes was for the first time studied in \cite{Hamed09} and reads
    \begin{equation}
    \label{4D_grass}
        \mathcal{L}_{n,k}=\int \frac{d^{k\times n}C}{\textrm{Vol}(GL(k))}\frac{1}{(12\cdots k)\ldots(n1\cdots k-1)}\delta^{(n-k)\times2}(\tilde{C}\lambda)\delta^{k\times2}(C\tilde{\lambda})\delta^{(n-k)\times4}(\tilde{C}\tilde{\eta})~,
    \end{equation}
where $\tilde{C}$ is the orthogonal complement to the Grassmannian $C$ satisfying $C\tilde{C}^T=0$ and $(1\ldots k)$ is a minor of consecutive columns from $1$ to $k$, etc.

The $k$ rows in the Grassmannian $k\times n$ matrix $C$ can be viewed as a $k$-plane in $n$ dimensions. Similarly, $\tilde{C}$ is an $(n-k)\times k$ matrix that can be viewed as an $(n-k)$-plane.

The massless $\mathcal{N}=4$ SYM theory enjoys both superconformal and dual superconformal symmetry. The corresponding 4D dilaton generator takes the form
\begin{equation}
    d_{4D}=\sum_{i}\left[\frac{1}{2}\lambda^{\alpha}_i\frac{\partial}{\partial \lambda^{\alpha}_i}+\frac{1}{2}\tilde{\lambda}^{\dot{\alpha}}_i\frac{\partial}{\partial \tilde{\lambda}^{\dot{\alpha}}_i}+1\right]~.
\end{equation}

\subsection{6D symplectic Grassmannian (via scattering equations)}

6D formulas based on the symplectic Grassmannian have been used recently to show the equivalence of two formulations for tree-level scattering of $n$ massless particles in 6D. In \cite{Schwarz} it was shown that a formula for scattering amplitudes based on rational maps \cite{Cachazo,Heydeman17,Heydeman18}, and another based on polarized scattering equations \cite{Geyer18}, are two different $GL(n, \mathbb{C})$ gauge fixings of a symplectic (Lagrangian) Grassmannian. The specific example was given by dimensional reduction of a 6D maximal SYM formula to obtain a formula for amplitudes of 4D massive $\mathcal{N}=4$ SYM on the Coulomb branch
\begin{equation}
    \mathcal{A}^{\mathcal{N}=4 CB}_n(\alpha)=\int d\mu^{\mathcal{N}=4 CB}_n \delta^{2\times n}(V \Omega  \eta^{I})\det{}^{\prime}H^{CB}_n~\mathrm{PT}(\alpha)~,
\end{equation}
where $V$ satisfy $V\Omega V^T=0$, PT($\alpha$) is the Parke-Taylor factor, the exact definition of $H$ can be found in \cite{Schwarz}
 and 
 \begin{equation}
     \int d\mu^{\mathcal{N}=4~CB}_n=\int \frac{d^n\sigma d^nv d^{2n}u}{\textrm{Vol}(SL(2,\mathbb{C}))_{\sigma}\times \textrm{Vol}(SL(2,\mathbb{C}))_{u}}\delta^{2\times n}(V \Omega  \Lambda^{\alpha})\delta^{2\times n}(V\Omega \tilde{\Lambda}^{\dot{\alpha}})~.
 \end{equation}

However, such formulation of amplitudes, although based on a symplectic Grassmannian, is still expressed in world-sheet coordinates. Hence a formulation using  Pl\"{u}cker coordinates is still missing.

\section{Symplectic Grassmaniann formula and symmetries}

As discussed in the previous section, the symplectic Grassmannian naturally talks to the kinematics of massless particles in 6D. In this section we fix some properties of the Grassmannian integral and discuss possible symmetries.
  
\subsection{Symplectic Grassmannian formula in 6D}

Based on the analogy with other Grassmannian formulas we propose that the 6D Grassmannian formula should contain $\delta$-functions relating the $C$-matrix and the $\Lambda$-matrix. For supersymmetric theories, the formula contains $\delta$-functions relating $C$-matrix and $\eta$-matrix, $\delta$-functions encoding the geometry of the Grassmannian (e.g.\ the symplecticness), a theory-dependent function of the $C$-matrix, and a measure factor, 
\begin{equation}
\label{proposal}
    \int \frac{d^{2n^2}C}{\textrm{Vol}(GL(n))} f^{6D}(C)\delta^{n\times 4}(C \Omega  \Lambda^T)\delta^{n\times\mathcal{N}}(C \Omega  \eta^T)\delta^{\frac{n(n-1)}{2}}(C \Omega  C^T)~,
\end{equation}
where $C\in \textrm{Gr}(n,2n)$ can be viewed as a $n\times2n$ matrix.

Now let us discuss the geometric "gauge fixing" represented by the $\textrm{Vol}(GL(n))$ factor in \eqref{proposal}. The $C$ as an element of $\textrm{Gr}(n,2n)$ describes a plane and can be viewed as a $n\times 2n$ matrix with $2n^2$ entries, or equivalently as $n$ row vectors in a $2n$ dimensional vector space. Those $n$ vectors span an $n$-dimensional plane. However, any non-degenerate linear transformation of those vectors span the same plane, in other words the same configuration $C$. The sought-for model therefore exhibits a non-compact $GL(n)$ symmetry that needs to be gauge fixed.
This is very similar to the gauge fixing in gauge theories, where gauge redundancy give rise to divergent integrals, which is also the case here. Naively without gauge fixing, the number of integrations of $C$-matrix entries in formula \eqref{proposal} is always greater than the number of $\delta$-functions $2n^2>4n+\frac{n(n-1)}{2}-6$, 
which gives the divergent integral. 
Therefore, we have to gauge fix the redundant degrees of freedom in $C$.

We can now calculate the number of integration left after gauge fixing 
\begin{equation}
    \frac{n^2-7n+12}{2}= \frac{(n-3)(n-4)}{2}~,
\end{equation}
where 6 $\delta$-functions were left for the momentum conservation. This imply that there will be no integration for $n=3$ and $n=4$, i.e.\ the first non-trivial integration appears at $n=5$. This should be compared to the 4D massless $\mathcal{N}=4$ SYM theory where the first integration appears at $n=6$. 

Let's now deduce the homogeneous $GL(n)$ weight of the function $f^{6D}(C)$  with respect to the $GL(n)$ scaling. The same plane (i.e.\ element in Grassmannian $\textrm{Gr}(n,2n)$) can be described by two matrices $C$ and $C^{\prime}$ related by a linear transformation
\begin{equation}
    C^{\prime}=L C, \qquad C,~C^{\prime}\in \textrm{Gr}(n,2n),~L\in GL(n)~.
\end{equation}
Individual parts of the Grassmannian formula transforms as
\begin{itemize}
    \item $\textrm{d}^{2n^2}C^{\prime}=\det^{2n}(L)~\textrm{d}^{2n^2}C$,
    
    \item $\delta^{4n}\left(C^{\prime}\Omega\Lambda^T\right)=\frac{1}{\det^4(L)}\delta^{4n}\left(C\Omega\Lambda^T\right)$,
    
    \item $\delta^{\mathcal{N}n}\left(C^{\prime}\Omega\eta^T\right)=\det^{\mathcal{N}}(L)\delta^{\mathcal{N}n}\left(C\Omega\eta^T\right)$,
    
    \item $\delta^{\frac{n(n-1)}{2}} \left( C^{\prime}\Omega C^{\prime T}\right) = \frac{1}{\det^{n-1}(L)} \delta^{\frac{n(n-1)}{2}} \left(C\Omega C^T\right)$.
\end{itemize}
The last equality is a bit more involved and therefore discussed in appendix \ref{Dirac-antisymm}. Consequently, if we demand the $GL(n)$ invariance of the integral, the function $f^{6D}(C)$ must scale like
\begin{equation}
\label{homogenity}
    f^{6D}(C^{\prime})=\det{}^{3-n-\mathcal{N}}(L)f^{6D}(C)~.
\end{equation}
It turns out to be convenient to construct the $GL(n)$-invariant integrand from manifestly $SL(n)$-invariant building blocks.
It is clear that if we restrict ourselves to $SL(n)$, it is sufficient to require  from the function $f(C)$ to be composed out of minors of $C$. A priori it is not clear why the "gauge fixing" group should be promoted to $GL(n)$, however, later in the end of subsection \ref{implications} we will show that this is equivalent to invariance under the dual dilaton generator $D$.

\subsection{6D superconformal symmetry and its breaking}
\label{5_2}
The formulas \eqref{3D_grass}, \eqref{4D_grass} and \eqref{proposal} are naturally (super)conformal invariant. This can be easily seen by applying the conformal dilaton generator to the Grassmannian formula in the appropriate dimension \eqref{3D_grass}, \eqref{4D_grass} and \eqref{proposal}, see Table~\ref{Dilatons}.  A similar argument holds for all other generators in the (super)conformal algebra. Such formulas contain external kinematical variables $\lambda$ in $\delta$-functions of the form $\delta(C\Lambda^T)$. By simple counting we find that these formulas are annihilated by the conformal dilaton generators in Table \ref{Dilatons}.

\begin{table}[]
    \centering
    \begin{tabular}{|c|c|}
    \hline
    Dimension &  Dilaton   \\\hline
   &\\
         3 &  $d=\sum_i\left[\frac{1}{2}\lambda_i^{\alpha}\frac{\partial}{\partial\lambda_{i}^{\alpha}}+\frac{1}{2}\right]$\\
         &\\
         \hline
         &\\
         4 &  $d=\sum_i\left[\frac{1}{2}\lambda_i^{\alpha}\frac{\partial}{\partial\lambda_{i}^{\alpha}}+\frac{1}{2}\tilde{\lambda}_i^{\dot{\alpha}}\frac{\partial}{\partial\tilde{\lambda}_i^{\dot{\alpha}}}+1\right]$\\ 
         &\\
         \hline
         &\\
         6 &$d=\sum_i\left[\frac{1}{2}\lambda^{Aa}_i\frac{\partial}{\partial\lambda^{Aa}_i}+2\right]$\\ &\\
         \hline
    \end{tabular}
    \caption{Dilatons in various dimensions \cite{Bargheer10, Plefka09, Huang_Lipstein10}. }
    \label{Dilatons}
\end{table}

This implies that non-conformal theories cannot be described by such simple Grassmannian formulas.
Let us break the conformal symmetry
by generalizing the function $f(C)$ to depend on external kinematical variables: 
\begin{equation}
\label{gen_funct}
    f(C) ~\rightarrow~f(C,\lambda)~.
\end{equation}
The reader may ponder if the function \eqref{gen_funct}  could  depend on $\eta$ also? This is not possible for 4D massive $\mathcal{N}=4$ SYM due to the $R$-symmetry generators of 6D $\mathcal{N}=(1,1)$ SYM. We have two $R$-symmetry generators \cite{Plefka14}
\begin{equation}
    b=\sum_{i=1}^n\left(\xi_{ia}\partial^a_i-1\right) \qquad\rightarrow \qquad \sum_{i=1}^n\left(\eta^{1a}_i\partial_{i1a}-1\right)~,
\end{equation}
\begin{equation*}
    \tilde{b}=\sum_{i=1}^n\left(\tilde{\xi}^{\dot{a}}_i\partial_{i\dot{a}}-1\right)\qquad \rightarrow\qquad \sum_{i=1}^n\left(\eta^{2a}_i\partial_{i2a}-1\right)~,
\end{equation*}
where on the left are the hypercharges written in the notation of \cite{Plefka14}, while on the right they are written in the chiral notation used in this paper.  We can easily write  the sum of $R$-symmetry generators in chiral language as $b+\tilde{b}=\sum_{i}\left(\eta^{Ia}_i\partial_{iIa}-2\right)$. Thus the total Grassmann degree grows as $2n$ and equally for $I=1,2$. This is completely captured by the Grassmann-odd $\delta$-functions
\begin{equation}
\label{ferm_delta_def}
    \delta^{2n}(C\Omega\eta^T)=\prod_{I=1}^2\prod_{m=1}^n\delta\left(\sum_{i=1}^nC^{ma}_i\epsilon_{ab}\eta^{Ib}_i\right)~.
\end{equation}
The ansatz for the Grassmannian formula having the symmetries of massive 4D $\mathcal{N}=4$ SYM  therefore becomes
\begin{equation}
\label{6D_formula}
    \int \frac{d^{2n^2}C}{\textrm{Vol}(GL(n))}f^{6D}(C,\lambda)\delta(C\Omega\Lambda^T)\delta(C\Omega C^T)\delta(C\Omega \eta^T)~,
\end{equation}
where in the rest of the paper we assume that the function $f^{6D}(C,\lambda)$ depends on the minors of the matrix $C$ and has a rational form with homogenity weights in minors given by eq.\ \eqref{homogenity}. A Grassmannian formula where the function $f$ 
depends also on $\lambda$, can be found e.g.\ in \cite{Lipstein20} for $\mathcal{N}=7$ SUGRA or in \cite{Heslop16} for $\mathcal{N}=8$ SUGRA.

\subsection{Little group invariance of an amplitude}

The task is now to restrict the function $f^{6D}(C,\lambda)$ by requiring pertinent symmetries. We start by imposing symmetries originating from kinematics, e.g.\ the little group. The superamplitudes of  massive 4D $\mathcal{N}=4$ SYM and its parent theory 6D $\mathcal{N}=(1,1)$ SYM are by construction invariant under the action of the corresponding little group, because the superfield is a scalar \cite{Plefka14}. Thus we demand little group invariance of formula \eqref{6D_formula}.

Let us briefly discuss the $n$-point little group. Spinors describing 6D massless momenta can be represented as $4\times 2$ matrices, i.e.\ 8 real (before complexification) degrees of freedom (DOF), however, the 6D on-shell momenta has 5 DOF only. The surplus is precisely removed by the 3 DOF of the 4D massive little group\footnote{To be precise, the massive little group of the double cover $Spin(3,1)$ of the Lorentz group.} $Spin(3)\cong SU(2)$. Therefore we should consider the total little group\footnote{The $USp(2n)$ is also sometimes called $Sp(n)$. To avoid confusion let us define $USp(2n):=Sp(2n, \mathbb{C})\cap U(2n)$}  to be  $\bigtimes_{i=1}^n SU(2) \subseteq USp(2n)$. The group $USp(2n)$ is, of course, much bigger than $n$ copies of $SU(2)$. The $Sp(2n)$ part is obvious from the $\delta$-function structure in eq.\ \eqref{6D_formula}, while the $U(n)$ part preserves the reality of the momenta (in case of real momenta).

We will use the "local" $SU(2)$ description when we use just part of the $C$ matrix, e.g.\ in minors, while we will use the "global" $USp(2n)$ description when the full $C$ matrix will be used, e.g. in $\delta(C\Omega \Lambda^T)$. The global picture can be used to deduce the little group properties of $f^{6D}(C,\lambda)$. Let us assume a little group transformation
\begin{equation}
     \textrm{M}_{2n\times2n}\in (U)Sp(2n)~,
\end{equation}
that acts on spinors in the following way
 \begin{equation}
     \Lambda^{\prime A}=\Lambda^{A} \textrm{M}~, \qquad \eta^{\prime I}=\eta^I \textrm{M}~.
 \end{equation}
This induces a transformation of $C$ matrix (using the relation $\textrm{M}\Omega\textrm{M}^T=\Omega$)
 \begin{equation}
     C^{\prime}=C\textrm{M}~,
 \end{equation} 
and the product of $\delta$-functions in eq.\ \eqref{6D_formula} becomes
\begin{equation}
    \delta(C^{\prime}\Omega \Lambda^T)\delta(C^{\prime}\Omega C^{\prime T})\delta(C^{\prime}\Omega \eta^T)~.
\end{equation}
Since the measure transforms as
\begin{equation}
    d^{2n^2}C^{\prime}=\det{}^n\textrm{M}~d^{2n^2}C~,
\end{equation}
and the determinant $\det\textrm{M}=1$ is unity for $\textrm{M}\in (U)Sp(2n)$, the measure is invariant. We conclude that the function $f^{6D}$ should be invariant under little group transformations
\begin{equation}
    f^{6D}(C^{\prime},\Lambda^{\prime})=f^{6D}(C,\Lambda)~.
\end{equation}

We will now find group invariant minors and combinations of minors that can serve as building blocks for the $f^{6D}$ function. To do this we use the "local" little group. Without loss of generality we may assume that the little group transforms the first particle only
\begin{equation}
    U^b_a=\left(
    \begin{array}{cc}
        \alpha_1 &\beta_1  \\
         \gamma_1& \delta_1
    \end{array}
    \right)~, \qquad U^b_a\in SU(2)~,
\end{equation}
\begin{equation}
    \lambda^A_{1a}=\left(
    \begin{array}{cc}
        \lambda^{A}_{11} &\lambda^{A}_{12}  
    \end{array}\right)_{4\times 2}~, \qquad 
     \lambda^{\prime A}_{1a}=\lambda^{A}_{1b}U^b_a~.
\end{equation}
Thus the $\Lambda$-plane  
\begin{equation}
    \Lambda^A{}_{\mu}~=~\left(\Lambda^A_{11}~\Lambda^A_{21} ~\Lambda^A_{31}~\ldots~ ~\Lambda^A_{n1}~ \Lambda^A_{12}~ ~\Lambda^A_{22}~\ldots~\Lambda^{A}_{n2}\right)_{4\times 2n}
\end{equation}
transforms with the matrix
\begin{equation}
    M^{\mu}_{\nu}=\left(
    \begin{array}{cccccc}
        \alpha_{1} &0&\ldots&\beta_{1}&\ldots &0  \\
         0& 1&\ldots&0&\ldots&0\\
         \vdots&\vdots&&\vdots&&\vdots\\
         \gamma_{1}&0&\ldots&\delta_{1}&\ldots&0\\
         \vdots&\vdots&&\vdots&&\vdots\\
         0&0&\ldots&0&\ldots&1
    \end{array}
    \right)_{2n\times 2n}~\in USp(2n)~.
\end{equation}
Let us for clarity consider a transformation on a 4-point minor (but the argument also holds for higher points). The minor containing the $1^{st}$ column of the $C$ matrix now transforms non-trivially with the little group
\begin{equation}
    \left(1~i~j~k\right)^{\prime}=\left(\alpha_1 c_{\bullet 1}+\gamma_1 c_{\bullet
    n+1}~i~j~k\right)=\alpha_1\left(1~i~j~k\right)+\gamma_1\left(n+1~i~j~k\right)
\end{equation}
where $i,j,k\neq 1$ and the dot on $c_{\bullet 1}$ denotes the row index. This suggests to include the $(n\!+\!1)$th column in the minor, which  is now little group invariant
\begin{equation}
    \left(1~n+1~j~k\right)^{\prime}=\left(\alpha_1 c_{\bullet 1}+\gamma_1 c_{\bullet n+1}~\beta_1 c_{\bullet 1}+\delta_1 c_{\bullet n+1}~j~k\right)=
\end{equation}
\begin{equation*}
    =\det\left((1~n+1~j~k)_{4\times4}\left(
    \begin{array}{cccc}
         \alpha_1&\beta_1&0&0 \\
         \gamma_1&\delta_1&0&0\\
         0&0&1&0\\
         0&0&0&1
    \end{array}
    \right)
    \right)=(1~n+1~j~k)~.
\end{equation*}
Thus the little group invariant and $GL(n)$ covariant objects are minors of the form:
\begin{equation}
(i~i+n ~j~j+n)~,   
\end{equation}
which can be written in a manifestly little group invariant way as
\begin{equation}
\label{minors_ansatz}
    (i~i+n~j~j+n)~\sim~(i^a~i_a~j^b~j_b)~, 
\end{equation}
where we use $C^{ma}_i$ with convention $C^{m1}_i:=C^m_{i}$ and $C^{m2}_i:=C^m_{i+n}$.

This immediately raises the question: How about odd particle number minors? Even number minors can be shown to be little group invariant if there appears both $i$th and $(i\!+\!n)$th column of the $C$ matrix in the minor. From massive 4D perspective the massive particles (massive vector $\mathcal{W}$-bosons) appear in pairs (\cite{Cachazo}, discussion at the end of p.\ 61). Therefore the last particle must be massless! In that case the little group reduces to just $U(1)$. Consequently it is enough to have same appearance of $i$th and $(i\!+\!n)$th column to cancel the little group scaling between multiplied minors. Can it be done also in 6D? The antisymmetric version for odd number of particles:
\begin{equation}
\label{minor_odd}
    (i~j~k~1^a)(1_a~l~m~n)~.
\end{equation}
This can be show to be little group invariant by direct calculation:
\begin{equation}
    (i~j~k~1)^{\prime}=\alpha_1(i~j~k~1)+\gamma_1(i~j~k~n+1)~,
\end{equation}
\begin{equation}
    (n+1~l~m~n)^{\prime}=\beta_1(1~l~m~n)+\delta_1(n+1~l~m~n)~.
\end{equation}
Plugging these transformation rules into eq.\ \eqref{minor_odd} proves the invariance. 

Let us mention for completeness that there is another possibility of little group invariants: the mix of minors and $\lambda$-spinors (or in principle $\eta$-spinor).

\section{Dual conformal symmetry of amplitudes and its implications}

In order to probe the Grassmannian formula we impose all symmetries of massive amplitudes, e.g. dual conformal symmetry \cite{Plefka14}. For more detailed discussions on dual conformal symmetries, see \cite{Drummond_Henn06, Brandhuber08, Caron-Huot10, Dennen11,Huang-Lipstein10,Gang11,Bhattacharya16}. An important part of dual conformal symmetry is the special dual conformal generator $K^{AD}$, which can be written with help of dual conformal inversion as
\begin{equation}
\label{K_inversion}
    K^{AD}:=I~P_{AD}~I~,
\end{equation}
where $P_{AD}$ is the dual translation generator. A deeper discussion of chiral 6D dual (super)conformal algebra can be found in \cite{6D_nase}. The action of special dual conformal generator can be derived from inversion properties of amplitudes. Well-known examples are 4D massless $\mathcal{N}=4$ SYM amplitudes\footnote{Our convention is that an amplitude $\mathcal{A}_n$ contains the (super)momentum conserving $\delta$-functions.} \cite{Drummond08}
\begin{equation}
    I[\mathcal{A}]=\prod_{i=1}^n\left(x_i^2\right)\mathcal{A}_n~,
\end{equation}
or 6D $\mathcal{N}=(1,1)$ SYM \cite{Plefka14}
\begin{equation}
\label{N(1,1)_weights}
    I[\mathcal{A}]=(x_1^2)^3\prod_{i=2}^n\left(x_i^2\right)\mathcal{A}_n~.
\end{equation}
A dimensional reduction of the latter leads to massive 4D amplitudes of $\mathcal{N}=4$ SYM on the Coulomb branch.

\subsection{6D dual conformal algebra and amplitudes with general weights}
In order to capture the dual conformal behaviour of amplitudes of various theories we consider the following weighted generalisation
\begin{equation}
\label{amplitude_general_weights}
    I[\mathcal{A}_n]=\prod_{i=1}^n\left(x_i^2\right)^{\alpha_i}\mathcal{A}_n~, \qquad \alpha_i\in\mathbb{R}~.
\end{equation}
Eq.\ \eqref{amplitude_general_weights} generalizes formula (85) in \cite{6D_nase}. This modification has direct consequences on the "covariance" under dual conformal generator $K^{AB}$. The generator \eqref{K_inversion} applied on the amplitude \eqref{amplitude_general_weights} (assuming invariance of the amplitude under $P_{AB}$) gives a non-trivial result
\begin{equation}
    K^{AB}\mathcal{A}_n=-\left(\sum_{i=1}^n\frac{\alpha_i}{2}x_i^{AB}\right)\mathcal{A}_n~,
\end{equation}
and thus is a priori not a symmetry of an amplitude. However, we can define a new generator
\begin{equation}
\label{K_prime}
    K^{\prime AB}:=K^{AB}+\sum_{i=1}^n\frac{\alpha_i}{2}x^{AB}_i~,
\end{equation}
which is again a symmetry of an amplitude \eqref{amplitude_general_weights}. In order to keep the conformal algebra, we have to modify the dual dilaton
\begin{equation}
\label{D_prime}
    D^{\prime}=D-\sum_{i=1}^n\alpha_i~.
\end{equation}
(This generalizes eqs.\ (85)-(91) in Ref.\ \cite{6D_nase}.) It has immediate consequences for the Grassmannian formula. Plugging the dual conformal weights of $\mathcal{N}=(1,1)$ SYM \eqref{N(1,1)_weights} into the dual dilaton \eqref{D_prime} we find
\begin{equation}
    D^{\prime}=D-n-2=-\sum_{i=1}^n\left[x_i^{AB}\partial_{iAB}+\frac{1}{2}\lambda^{Aa}_i\partial_{iAa}+\frac{1}{2}\theta^{Ia}_i\partial_{iIa}\right]-n-2~.
\end{equation}
Let's now transfer the dual dilaton from dual superspace to the on-shell superspace. This is motivated by Ref.\ \cite{Plefka09}. We can drop the terms with derivatives acting on dual variables, so the dual dilaton now takes the form
\begin{equation}
\label{dilaton_onshell}
     D^{\prime}|_{\textrm{on-shell}}=-\sum_{i=1}^n\left[\frac{1}{2}\lambda^{Aa}_i\partial_{iAa}\right]-n-2~,
\end{equation}
which agree with the symmetry generator of 6D $\mathcal{N}=(1,1)$ SYM (cf.\ eq.\ (4.20) in \cite{Plefka14}).

The special dual conformal generator \eqref{K_inversion} in the dual chiral super-space has the explicit form \cite{6D_nase}
\begin{equation}
\label{K_6D}
K^{AD}=\frac{1}{2}\sum_i   \left\{x_i^{[AE}\theta^{MD]}_i\partial_{iME}-x^{[AB}_i x^{CD]}_i\partial_{iBC}+\theta^{M[A}_i\theta^{ND]}_i\partial_{iMN} \right.
\end{equation}
\begin{equation*}
~~~~~~~~~~+\frac{1}{2}\left(x^{[AE}_i+ x^{[AE}_{i+1}\right)\lambda^{D]a}_i\partial_{iEa}+\left.\frac{1}{2}\lambda_i^{[Aa}\left(\theta_{i}^{MD]}+\theta_{i+1}^{MD]}\right)\partial_{iMa}\right\}~.
\end{equation*}
Let us bring this generator to the on-shell superspace. Similarly to the dilaton, we can remove terms containing derivatives w.r.t.\ dual coordinates 
\begin{equation}
    K^{AD}|_{\textrm{on-shell}}=\frac{1}{4}\sum_i \left\{\left(x^{[AE}_i+ x^{[AE}_{i+1}\right)\lambda^{D]a}_i\partial_{iEa}+\lambda_i^{[Aa}\left(\theta_{i}^{MD]}+\theta_{i+1}^{MD]}\right)\partial_{iMa}\right\}~.
\end{equation}
We next express all dual variables as  functions of $x_1$, $\theta_1$, $\lambda_i$ and $\eta_i$ using the telescopic sum solution to the dual constraints
\begin{equation}
\label{constraint_solution}
    x_i^{AB}=x_1^{AB}-\sum_{j<i}\lambda^{Aa}_j\lambda^{B}_{ja}, \qquad \theta^{IA}_i=\theta^{IA}_1-\sum_{j<i}\eta^{Ia}_j\lambda^{A}_{ja}~.
\end{equation}
We see that there will be some terms proportional to the $x_1^{AB}$ or $\theta_1^{IA}$
\begin{equation}
\label{what_remains}
    \frac{1}{2}\sum_i\left\{x^{[AE}_1\lambda^{D]a}_i\partial_{iEa}+\lambda^{[Aa}_i\theta^{MD]}_1\partial_{iMa}\right\}=
\end{equation}
\begin{equation*}
    = x^{[AE}_1\bar{m}^{D]}_E-\frac{1}{2}x^{AD}_1D|_{\textrm{on-shell}}+\frac{1}{2}\theta^{M[D}_1\tilde{q}^{A]}_M~.
\end{equation*}
Inserting eq.\ \eqref{what_remains} into the modified $K^{\prime}$ we get the "1-part" to be
\begin{equation}
    x^{[AE}_1\bar{m}^{D]}_E-\frac{1}{2}x^{AD}_1D^{\prime}|_{\textrm{on-shell}}+\frac{1}{2}\theta^{M[D}_1\tilde{q}^{A]}_M~.
\end{equation}
If we now assume that $f^{6D}(C,\lambda)$ is Lorentz invariant, i.e.\ is annihilated by the $\bar{m}^D_{E}$, and all the generators $\bar{m}$, $D^{\prime}$ and $\tilde{q}$ annihilate formula \eqref{6D_formula}, then we can conclude that all terms proportional to $x_1$ or $\theta_1$ annihilate eq.\ \eqref{6D_formula} and therefore can be neglected. What remains is
\begin{equation}
\label{K_onshell_middle}
    -\frac{1}{4}\sum_i\left(\sum_{j<i}+\sum_{j<i+1}\right)\lambda^{[Ab}_{j}\lambda^{E}_{jb}\lambda^{D]a}_i\partial_{iEa}
    -\frac{1}{4}\sum_i\left(\sum_{j<i}+\sum_{j<i+1}\right)\lambda^{[Aa}_i\eta^{Mb}_{j}\lambda^{D]}_{jb}\partial_{iMa}~.
\end{equation}
We can now add to eq.\ \eqref{K_onshell_middle} the expression
\begin{equation}
\label{compensator}
    \frac{1}{4}q^{M[D}\bar{q}^{A]}_M+\frac{1}{4}p^{[AE}\bar{m}^{D]}_E+\frac{1}{8}p^{AD}\sum_i\lambda^{Ba}_i\partial_{iBa}
\end{equation}
and use the relation
\begin{equation}
    \sum_{j<i}a_ib_j-\frac{1}{2}\left(\sum_i a_i\right)\left(\sum_{j}b_j\right)=\frac{1}{2}\sum_{j<i}(a_ib_j-(i\leftrightarrow j))-\frac{1}{2}\sum_{i}a_ib_i~.
\end{equation}
First two terms in \eqref{compensator} annihilate the amplitudes manifestly. The last term contains the Euler ("counting") operator, and therefore its action on the amplitude is proportional to the amplitude itself.  Consequently the last term in \eqref{compensator} is proportional to $p^{AD}$ and annihilate the amplitude. 
The result is in a bi-local form
\begin{equation}
\label{pomocny_vypocet}
\begin{aligned}
    -&\frac{1}{4}\sum_{j<i}\left(
    \lambda^{[Ab}_{j}\lambda^{E}_{jb}\lambda^{D]a}_i\partial_{iEa}+\lambda^{[Aa}_i\eta^{Mb}_{j}\lambda^{D]}_{jb}\partial_{iMa}
    -(i\leftrightarrow j)\right)=\\
    -&\frac{1}{4}\sum_{j<i}\left(p^{[AE}_jm_{iE}^{D]}+q^{M[D}_j\bar{q}^{A]}_{iM}-(i\leftrightarrow j)\right)~.
\end{aligned}
\end{equation}
The same modification can be done with the additional part in \eqref{K_prime} and we get the so-called bi-local formula \cite{Plefka09, Huang-Lipstein10}
\begin{equation}
\label{K_onshell}
    K^{\prime AD}_{\textrm{on-shell}}=-\frac{1}{4}\sum_{j<i}\left(p^{[AE}_j\bar{m}^{D]}_{iE}-p^{AD}_jD^{\prime}_{i|\textrm{on-shell}}+q^{M[D}_j\bar{q}^{A]}_{iM} -(i\leftrightarrow j)\right)+\frac{1}{4}\sum_i\alpha_ip^{AD}_i~,
\end{equation}
where $D^{\prime}_{i|\textrm{on-shell}}=-\frac{1}{2}\lambda^{Aa}_i\partial_{iAa}-\alpha_i$. Due to the last term, the form \eqref{K_onshell} is very similar to the so-called evaluation representation of the Yangian algebra in \cite{Ferro14}, eq.\ (10). This might be relevant for the construction of amplitudes in theories with potential Yangian symmetry and dual conformal weights different from those of 4D massless $\mathcal{N}=4$ SYM.

Although the bi-local form \eqref{K_onshell} is often useful for proofs regarding Yangian symmetry \cite{Huang-Lipstein10, Plefka09}, we shall here use a different form of it. We can rewrite eq.\ \eqref{pomocny_vypocet} in the form
\begin{equation}
\begin{aligned}
\label{K-1}
    K^{AD}|_{\textrm{on-shell}}=&-\frac{1}{4}\left(\sum_{j<i}-\sum_{j>i}\right)\left[\lambda^{[Ab}_{j}\lambda^{E}_{jb}\lambda^{D]a}_i\partial_{iEa}+\lambda^{[Aa}_i\eta^{Mb}_{j}\lambda^{D]}_{jb}\partial_{iMa}\right]\\
    =&-\frac{1}{4}\left(\sum_{j<i}-\sum_{j>i}\right)\left[\lambda^{[Ab}_{j}\lambda^{D]a}_i\lambda^{E}_{jb}\partial_{iEa}+\lambda^{[Ab}_j\lambda^{D]a}_{i}\eta^{M}_{jb}\partial_{iMa}\right]\\
     =&-\frac{1}{4}\left(\sum_{j<i}-\sum_{j>i}\right)\lambda^{[Ab}_{j}\lambda^{D]a}_i\Lambda^{\mathcal{A}}_{jb}\partial_{ia\mathcal{A}}~,
\end{aligned}
\end{equation}
where the caligraphic index $\mathcal{A}$ denotes a superized index \eqref{superize} introduced in \cite{6D_nase}.  We skip the subscript "on-shell" from now on, because discussions in the  remaining chapters will encompass on-shell superspace only. The special dual conformal generator can be further simplified to
\begin{equation}
\label{K_useful}
    K^{AD}=-\frac{1}{4}\sum_{j<i}\Lambda^{[Ab}_j\Lambda^{D]a}_i\left(\epsilon_{ac}\Lambda^{\mathcal{A}}_{jb}\frac{\partial}{\partial \Lambda^{\mathcal{A}}_{ic}}+\epsilon_{bc}\Lambda^{\mathcal{A}}_{ia}\frac{\partial}{\partial \Lambda^{\mathcal{A}}_{jc}}\right)=-\frac{1}{4}\sum_{j<i}\Lambda^{[Ab}_j\Lambda^{D]a}_i\left(\mathcal{O}_{jb|ia}+\mathcal{O}_{ia|jb}\right)~,
\end{equation}
where we have defined the operator
\begin{equation}
    \mathcal{O}_{ia|jb}:=\epsilon_{bc}\Lambda^{\mathcal{A}}_{ia}\frac{\partial}{\partial \Lambda^{\mathcal{A}}_{jc}}~.
\end{equation}

\subsection{Implications of dual conformal algebra on Grassmannian formula}
\label{implications}
In the previous subsection we derived the modified generators of the dual conformal algebra that annihilate the amplitude \eqref{amplitude_general_weights}.
Let us find out what consequences this has for the Grassmannian formula \eqref{6D_formula}. The main idea is to write the amplitude with general weights \eqref{amplitude_general_weights} as an integral over the Grassmannian \eqref{6D_formula}. Moreover, we demand that the "obvious" symmetries of the amplitudes, such as momentum, Lorentz and little group, are manifest. This implies that the function $f^{6D}(C,\lambda)$ must contain $\lambda$ in Lorentz invariant combinations. Although there could exist complicated invariants mixing $C$ and $\lambda$, we assume that the function $f^{6D}(C,\lambda)$ factorizes
\begin{equation}
f^{6D}(C,\lambda)\quad \rightarrow \quad f^{6D}(C)~g^{6D}(\lambda)~.   
\end{equation}
The first non-trivial generator in the dual conformal algebra is the dual dilaton \eqref{dilaton_onshell}. Applying it to the Grassmannian formula \eqref{6D_formula} we find that the function $g^{6D}(\lambda)$ must scale in $\lambda$ with a power
\begin{equation}
\label{dilaton_4_weight}
    \#_{\lambda}g^{6D}(\lambda)=2(n-2)~,
\end{equation}
where $\#_{\lambda}$ denotes the power of $\lambda$ in $g^{6D}(\lambda)$. Therefore, we are able to construct amplitudes with even number of external particles only. (In fact, the Lorentz invariants in chiral 6D must contain multiples of 4 $\lambda$s.) The first non-trivial example is $n=4$ particles, where the function $g^{6D}(\lambda)$ contain 4 $\lambda$s. After imposing Lorentz and little group invariance, examples of $\lambda$-building blocks are the Mandelstam variables $s_{ij}\sim \langle i^ai_aj^bj_b\rangle$.

The last non-trivial generator of the dual conformal algebra is the special dual conformal generator $K^{\prime AD}$. The generator consists of two parts, the first part $K^{AB}$ is a first-order linear differential operator, and the second part is an inhomogeneous part, see eq.\ \eqref{K_prime}. Let us now focus on the first part $K^{AB}$. When acting with $K^{AD}$ on the Grassmannian formula \eqref{6D_formula}, we use Leibnitz rule 
\begin{equation}
\label{K_on_grass}
\begin{aligned}    
&{K^{AD}\int \frac{d^{2n^2}C}{\textrm{Vol}(GL(n))}g^{6D}(\lambda)f^{6D}(C)\delta(C\Omega\Lambda^T)\delta(C\Omega C^T)\delta(C\Omega \eta^T)~=}\\
&K^{AD}\left[g^{6D}(\lambda)\right]\int \frac{d^{2n^2}C}{\textrm{Vol}(GL(n))}f^{6D}(C)\delta(C\Omega\Lambda^T)\delta(C\Omega C^T)\delta(C\Omega \eta^T)~\\
&+g^{6D}(\lambda)\int \frac{d^{2n^2}C}{\textrm{Vol}(GL(n))}f^{6D}(C)\delta(C\Omega C^T)K^{AD}\left[\delta(C\Omega\Lambda^T)\delta(C\Omega \eta^T)\right]~.
\end{aligned}
\end{equation}
Since  $g^{6D}(\lambda)$ must be Lorentz invariant, we assume that it depends on angle brackets \eqref{6D_anglebracket}. Therefore, it is useful to find the action of $K^{AD}$ on angle-brackets (via the chain rule). Without loss of generality we can order the angle bracket entries such that $l<m$. We find after some algebra manipulations  
\begin{equation}
\begin{aligned}
\label{K_on_angle}
    K^{AD}<l^a~l_a~m^b~m_b>=&-\frac{1}{2}\left(\sum_{j<l}-\sum_{j>m}\right)p_j^{AD}<l^a~l_a~m^b~m_b>\\
    &-\frac{1}{2}\sum_{m>j>l}\left(p_m^{AD}<j^a~j_a~l^b~l_b>-p^{AD}_l<j^a~j_a~m^b~m_b>\right)~.
\end{aligned}
\end{equation}
For detailed calculation, see Appendix \ref{app_K_on_angle}.

The second term in eq.\ \eqref{K_on_grass} deserves a deeper discussion. We would like to rewrite this equation s.t.\ the generator acts on the unknown function $f^{6D}(C)$. This can be done by replacing
\begin{equation}
    \Lambda^{\mathcal{A}}_{jb}\frac{\partial}{\partial \Lambda^{\mathcal{A}}_{ic}}\qquad \rightarrow \qquad -\sum_{p=1}^nC^{pc}_i\frac{\partial}{\partial C^{pb}_j}~,
    \label{lambdacreplacement}
\end{equation}
and integrate by parts. We see that the generator $K^{AD}$ now depends on $C$ and hits the product $f^{6D}(C)\delta(C\Omega C^T)$. The $\delta$-function $\delta(C\Omega C^T)$ is, however, annihilated by this generator. Schematically, this becomes
\begin{equation}
\label{helpargu}
    \sum_{p=1}^n\left(\epsilon_{ac}C^{pc}_i\frac{\partial}{\partial C^{pb}_j}+\epsilon_{bc}C^{pc}_j\frac{\partial}{\partial C^{pa}_i}\right)\left[\sum_{k=1}^nC^{re}_k\epsilon_{ef}C^{sf}_k\right]~=~0~,
\end{equation}
when we act on the $\delta$-function argument, i.e.\ the above square bracket, because
\begin{itemize}
    \item 
    \begin{equation}
         \sum_{p=1}^n\left(\epsilon_{ac}C^{pc}_i\frac{\partial}{\partial C^{pb}_j}\right)\left[\sum_{k=1}^nC^{re}_k\epsilon_{ef}C^{sf}_k\right]~=~C^r_{ia}C^{s}_{jb}-C^s_{ia}C^{r}_{jb}~,
    \end{equation}
    
    \item
    \begin{equation}
         \sum_{p=1}^n\left(\epsilon_{bc}C^{pc}_j\frac{\partial}{\partial C^{pa}_i}\right)\left[\sum_{k=1}^nC^{re}_k\epsilon_{ef}C^{sf}_k\right]~=~C^s_{ia}C^{r}_{jb}-C^r_{ia}C^{s}_{jb}~.
    \end{equation}
\end{itemize}

We assume that the function $f^{6D}(C)$ is a function of minors of the matrix $C$. Similar to the function $g^{6D}(\lambda)$, we have to investigate the action of $K^{AD}$ on minors of the $C$-matrix. Let us first define the following shorthand notation
\begin{equation}
\label{minor_def}
    (ij\ldots l):=(i^ai_bj^aj_b\ldots l^cl_c):=\epsilon_{mnpq\ldots rs}C^{ma}_iC^{n}_{ia}C^{pb}_jC^{q}_{jb}\ldots C^{rc}_lC^s_{lc}~.
\end{equation}
The action of the generator $K^{AD}$ on the minors relevant for the $n=4$ particle example is
\begin{equation}
\label{kadxy}
    K^{AD}(xy)=-\frac{1}{2}\left(\sum_{i>y}-\sum_{i<x}\right)p_i^{AD}(xy)-\frac{1}{2}\sum_{x<i<y}\left[p_x^{AD}(iy)-p_{y}^{AD}(ix)\right]~.
\end{equation}
The details of the calculation \eqref{kadxy} can be found in Appendix \ref{K_on_minor}, where both the gauge fixed and the more involved non-gauge fixed versions are listed.

It is interesting at this point to revisit the role of the dual dilaton generator. As we have already pointed out in previous paragraphs, the dual dilaton is a symmetry of amplitudes and determines the $\lambda$-weight of the function $g(\lambda)$. Therefore, what remains to act on the $\delta$-functions in the Grassmannian integral is the ordinary 6D  conformal dilaton (see Table~\ref{Dilatons}),
\begin{equation}
\label{dilaton_pomoc}
    \int \frac{{d}C}{\textrm{Vol}(GL(n))}f^{6D}(C)\delta^{\frac{n(n-1)}{2}}(C\Omega C^T)\delta^{n\mathcal{N}}(C\Omega\eta^T)\left[-\frac{1}{2}\sum_{i=1}^{n}\lambda^{Aa}_i\frac{\partial}{\partial \lambda^{Aa}_i}-2n\right]\delta^{4n}(C\Omega\lambda^T)=0~.
\end{equation}
Let us now replace the $\lambda\frac{\partial}{\partial \lambda}$ term according to \eqref{lambdacreplacement}
\begin{equation}
    \lambda^{Aa}_i\frac{\partial}{\partial \lambda^{Aa}_i}\qquad \rightarrow\qquad \sum_{p=1}^nC^{pb}_i\frac{\partial}{\partial C^{pb}_i}~,
\end{equation}
which is now again only the Euler counting operator in $C$. Next assume that boundary terms vanish in the following relation
\begin{equation}
\label{useful_zero}
    \int \frac{\textrm{d}C}{\textrm{Vol}(GL(n))}\frac{1}{2}\sum_{i,p,c}\frac{\partial}{\partial C^{pc}_i}\left[C^{pc}_if^{6D}(C)\delta(C\Omega C^T)\delta(C\Omega \eta^T)\delta(C\Omega\lambda^T)\right]~=~0~.
\end{equation}
The eqs.\ \eqref{dilaton_pomoc} and \eqref{useful_zero} now imply
\begin{equation}
\#_{C}f^{6D}(C)=n(3-n-\mathcal{N})~,
\end{equation}
where $\#_{C}f^{6D}(C)$ means the $C$-weight of $f^{6D}(C)$. If we assume that $f^{6D}(C)$ consists of minors of $C$ (each minor is of degree $n$), we get the scaling condition \eqref{homogenity}.  In other words, we have found that the 6D conformal dilaton invariance is equivalent to $GL(n)$ invariance of the Grassmannian formula.

\section{An example: 4-point amplitude for massive 4D $\mathcal{N}=4$ SYM on Coulomb Branch}
\label{example}
We have seen in subsection \ref{implications} that the dual dilaton implies the presence of a function of weight $2(n-2)$ in the  $\lambda$s. Thus  we can write an ansatz for the Grassmannian representation of an 4-pt. amplitude
\begin{equation}
    \mathcal{A}_4:=\sum_{j<i}\gamma_{ij}I_{ij}~, \qquad \gamma_{ij}\in \mathbb{C}~,
\end{equation}
where the basic building blocks $I_{ij}$ are
\begin{equation}
\label{4-p_ansatz}
    I_{ij}:=\langle i^a~i_a~j^b~j_b\rangle \int \frac{dC}{\textrm{Vol}(GL(n))}f_{ij}(C)\delta(C\Omega C^T)\delta(C\Omega \Lambda^T)~.
\end{equation}
The $f_{ij}(C)$ is the function that we want to determine. According to eq.\ \eqref{homogenity} it must have a weight $-3$ in minors \eqref{minors_ansatz}. We assume a rational ansatz of the form
\begin{equation}
\label{f_ansatz}
\begin{aligned}
     f_{ij}(C)=(12)^{a_{ij}}(23)^{b_{ij}}&(34)^{c_{ij}}(13)^{d_{ij}}(14)^{e_{ij}}(24)^{f_{ij}}~,  \\
 a_{ij},~ b_{ij},~ c_{ij},~ d_{ij},~e_{ij},~ f_{ij}\in \mathbb{R}~,& \qquad a_{ij}+ b_{ij}+ c_{ij}+ d_{ij}+e_{ij}+ f_{ij}=-3~.
\end{aligned}
\end{equation}

Next we use the fact that the dual conformal generator $K^{\prime AD}$ annihilates the amplitude ansatz
\begin{equation}
\label{K-equations}
    K^{\prime AD}\mathcal{A}_4=0~,
\end{equation}
which can be written in the form (omitting integral and $\delta$-functions)\footnote{For clarity we define $\llangle ij\rrangle:=\langle i^a i_a j^b j_b\rangle$.}
\begin{equation}
\label{4-p_equation}
   \sum_{j<i}\gamma_{ij}\left[(K^{AD}\llangle ij \rrangle )f_{ij}(C)-\llangle ij\rrangle (K^{AD}f_{ij}(C)) -\frac{1}{2}[3p^{AD}_1+2p^{AD}_2+p^{AD}_3]\llangle ij\rrangle f_{ij}(C)\right]=0~.
\end{equation}
The above eq.\ \eqref{4-p_equation}, however, is not unique.  We can add any multiple of the total momentum, because each basic building block contains a momentum-conserving $\delta$-function. Therefore the right hand side of eq.\ \eqref{4-p_equation} can be replaced by
\begin{equation}
\label{4-p_equation_II}
    \sum_{j<i}\gamma_{ij}l_{ij}(p^{AD}_1+p^{AD}_2+p^{AD}_3+p^{AD}_4)\llangle ij\rrangle f_{ij}(C)~,
\end{equation}
where $l_{ij}\in\mathbb{R}$ are Lagrange multipliers. (Both the above expressions are implicitly meant to be inside the Grassmannian integral over $C$.) 

In the first term of eq.\ \eqref{4-p_equation} we need to find the action of $K^{AD}$ on the angle bracket $\llangle ij\rrangle$. We use eq.\ \eqref{K_on_angle} and find
\begin{equation}
\label{K_on_angle_individual}
\begin{aligned}
    K^{AD}\llangle12\rrangle~=~&\frac{1}{2}(p_3^{AD}+p^{AD}_4)\llangle12\rrangle~,\\
    K^{AD}\llangle23\rrangle~=~&\frac{1}{2}(p^{AD}_4-p_1^{AD})\llangle23\rrangle~,\\
    K^{AD}\llangle34\rrangle~=~&\frac{1}{2}(p^{AD}_3+p^{AD}_4)\llangle12\rrangle~,\\
    K^{AD}\llangle13\rrangle~=~&-\frac{1}{2}\left(p_3^{AD}+p_4^{AD}\right)\llangle12\rrangle-\frac{1}{2}\left(p_4^{AD}-p_{1}^{AD}\right)\llangle23\rrangle~,\\
    K^{AD}\llangle14\rrangle~=~&\frac{1}{2}(p^{AD}_4-p^{AD}_1)\llangle23\rrangle~,\\
    K^{AD}\llangle24\rrangle~=~&-\frac{1}{2}\left(p_3^{AD}+p_4^{AD}\right)\llangle12\rrangle-\frac{1}{2}\left(p_4^{AD}-p_{1}^{AD}\right)\llangle23\rrangle~.
\end{aligned}
\end{equation}
However not all of the angle brackets are independent.  They are proportional to the  Mandelstam variables $s_{ij}$. We are allowed to use momentum conservation and thus 
\begin{equation}
    \llangle34\rrangle=\llangle12\rrangle\sim s_{12}~, \qquad \llangle24\rrangle=\llangle13\rrangle\sim s_{13}~, \qquad
    \llangle23\rrangle=\llangle14\rrangle\sim s_{14}~.
\end{equation}
Furthermore the Mandelstams for massless particles (and angle brackets also) satisfy the following relation
\begin{equation}
    s_{12}+s_{13}+s_{14}=0 \quad \Rightarrow\quad \llangle12\rrangle+\llangle13\rrangle+\llangle14\rrangle=0~.
\end{equation}
We choose $\llangle12\rrangle$ and $\llangle23\rrangle$ to be the independent variables.

The action of $K^{AD}$ on minors goes as follows
\begin{equation}
    K^{AD}(xy)=-\frac{1}{2}\left(\sum_{i>y}-\sum_{i<x}\right)p_i^{AD}(xy)-\frac{1}{2}\sum_{x<i<y}\left[p_x^{AD}(iy)-p_{y}^{AD}(ix)\right]~, 
\end{equation}
\begin{equation}
\label{K_on_minor_individual}
\begin{aligned}
    K^{AD}(12)~=~&-\frac{1}{2}(p^{AD}_3+p^{AD}_4)(12)~, \\
    K^{AD}(23)~=~&-\frac{1}{2}(p_{4}^{AD}-p_1^{AD})(23)~, \\
    K^{AD}(34)~=~&\frac{1}{2}(p_1^{AD}+p_2^{AD})(34)~, \\
    K^{AD}(13)~=~&-\frac{1}{2}p_4^{AD}(13)-\frac{1}{2}\left[p_1^{AD}(23)-p_3^{AD}(12)\right]~, \\
    K^{AD}(14)~=~&-\frac{1}{2}\left[p_1^{AD}(24)+p_1^{AD}(34)-p_4^{AD}(12)-p_4^{AD}(13)\right]~, \\
    K^{AD}(24)~=~&\frac{1}{2}p_1^{AD}(24)-\frac{1}{2}\left[p_2^{AD}(34)-p_4^{AD}(23)\right]~.
\end{aligned}
\end{equation}
We can now plug the previous results into  eqs.\ \eqref{4-p_equation} and \eqref{4-p_equation_II} and split them into 2 pieces according to the 2 independent Mandelstams. The $\llangle12\rrangle$ sector of eq.\ \eqref{4-p_equation} takes the form
\begin{equation}
\label{12sector}
\begin{aligned}
    &\frac{1}{2}\left[\gamma_{12}f_{12}(C)+\gamma_{34}f_{34}(C)\right]\left(p_3^{AD}+p_4^{AD}\right)-\frac{1}{2}\left[\gamma_{13}f_{13}(C)+\gamma_{24}f_{24}(C)\right]\left(p_3^{AD}+p_4^{AD}\right)\\
    &-\gamma_{12}\left[K^{AD}f_{12}(C)\right]-\gamma_{34}\left[K^{AD}f_{34}(C)\right]+\gamma_{13}\left[K^{AD}f_{13}(C)\right]+\gamma_{24}\left[K^{AD}f_{24}(C)\right]\\
    &-\frac{1}{2}\left[3p_1^{AD}+2p_2^{AD}+p_3^{AD}\right]\left(\gamma_{12}f_{12}(C)+\gamma_{34}f_{34}(C)-\gamma_{13}f_{13}(C)-\gamma_{24}f_{24}(C)\right)=\\
    &=(p_1^{AD}+p_{2}^{AD}+p_3^{AD}+p_4^{AD})
    \left[\gamma_{12}l_{12}f_{12}(C)+\gamma_{34}l_{34}f_{34}(C)-\gamma_{13}l_{13}f_{13}(C)-\gamma_{24}l_{24}f_{24}(C)\right]~,
\end{aligned}
\end{equation}
and the $\llangle 23\rrangle$ sector is
\begin{equation}
\label{23sector}
\begin{aligned}
    &\frac{1}{2}[\gamma_{23}f_{23}(C)+\gamma_{14}f_{14}(C)-\gamma_{13}f_{13}(C)-\gamma_{24}f_{24}(C)](p_4^{AD}-p_1^{AD})\\
    &-\gamma_{23}[K^{AD}f_{23}(C)]-\gamma_{14}[K^{AD}f_{14}(C)]+\gamma_{13}[K^{AD}f_{13}(C)]+\gamma_{24}[K^{AD}f_{24}(C)]\\
    &-\frac{1}{2}(3p^{AD}_1+2p_2^{AD}+p_3^{AD})[\gamma_{23}f_{23}(C)+\gamma_{14}f_{14}(C)-\gamma_{13}f_{13}(C)-\gamma_{24}f_{24}(C)]\\
    &=(p_1^{AD}+p_2^{AD}+p_3^{AD}+p_4^{AD})[\gamma_{23}l_{23}f_{23}(C)+\gamma_{14}l_{14}f_{14}(C)-\gamma_{13}l_{13}f_{13}(C)-\gamma_{24}l_{24}f_{24}(C)]~.
\end{aligned}
\end{equation}
Using the explicit action of $K^{AD}$ on minors
\begin{equation}
\begin{aligned}
    K^{AD}f_{ij}(C)=&\frac{f_{ij}(C)}{2}\left[-(p_3^{AD}+p_4^{AD})a_{ij}-(p^{AD}_4-p^{AD}_1)b_{ij}+(p_1^{AD}+p_2^{AD})c_{ij}-p_4^{AD}d_{ij}\right.\\
    &+p_1^{AD}f_{ij}-d_{ij}\frac{p_1^{AD}(23)-p_3^{AD}(12)}{(13)}-f_{ij}\frac{p_2^{AD}(34)-p_4^{AD}(23)}{(24)}\\
    &\left.-e_{ij}\left(p_1^{AD}\frac{(24)+(34)}{(14)}-p_4^{AD}\frac{(12)-(13)}{(14)}\right)\right]~,
\end{aligned}
\end{equation}
we can see that eqs.\ \eqref{12sector} and \eqref{23sector} separate according to the indeterminates $p_i^{AD}$, because the Lagrange multipliers imply that we can effectively treat the momentum variables $p_i^{AD}$ as independent (for fixed indices $A$ and $D$). These equations can be found in Appendix \ref{n=4_details}. Furthermore, we split according to the independent $f_{ij}(C)$ and solve the corresponding linear equations. The solution takes the form
\begin{equation}
    \begin{aligned}
        &f_{12}(C)=\left(\frac{(12)}{(34)}\right)^{2l_{12}}\left(\frac{(14)}{(23)}\right)^{e_{12}}\frac{1}{(23)(34)^2}~,~ f_{34}(C)=\left(\frac{(12)}{(34)}\right)^{2l_{34}}\left(\frac{(14)}{(23)}\right)^{e_{34}}\frac{1}{(23)(34)^2}~,\\
        &f_{23}(C)=\left(\frac{(12)}{(34)}\right)^{2l_{23}}\left(\frac{(14)}{(23)}\right)^{e_{23}}\frac{(12)}{(23)^2(34)^2}~,~
         f_{14}(C)=\left(\frac{(12)}{(34)}\right)^{2l_{14}}\left(\frac{(14)}{(23)}\right)^{e_{14}}\frac{(12)}{(23)^2(34)^2}~,\\
         &\gamma_{13}=\gamma_{24}=0~,
    \end{aligned}
\end{equation}
where $l_{ij},e_{ij}\in\mathbb{R}$.

\section{Exact integration of the 4-point example}
The spinor $\Lambda^I_{\mu}$ can be viewed as a pair of two square matrices
\begin{equation}
    \Lambda^{I}{}_{\mu}=(A_{4\times4}~B_{4\times4})_{4\times8}~.
\end{equation}
The $C$ matrix can be gauge fixed to the following form
\begin{equation}
\label{cad}
    C^m{}_{\mu}=(A_{4\times4}~D_{4\times4})_{4\times8}~.
\end{equation}
This gauge fixing \eqref{cad} only makes sense for $n=4$, because the dimensions of the $C$ and $\Lambda$ matrices then match. 
The $\delta$-function equations now become
\begin{equation}
   \delta(C\Omega\Lambda):~~ DA^T=AB^T~, \qquad \delta(C\Omega C^T):~~DA^T=AD^T~.
\end{equation}
The solution of $\delta(C\Omega\Lambda)$ on the support of $\delta(C\Omega C^T)$ can be rewritten
\begin{equation}
    AD^T=AB^T, ~\qquad \Rightarrow \qquad D=B~,\qquad \Rightarrow\qquad C=\Lambda~.
\end{equation}
  The $C$-matrix now becomes the $\Lambda$-matrix, consequently we can write\footnote{In steps denoted by "$\rightarrow$" we ignore proportionality constants, because they can be absorbed into $\gamma_{ij}$.}
\begin{equation}
   (ij) \rightarrow \llangle ij\rrangle~,
 \end{equation}
which for the $f_{ij}(C)$ imply
\begin{equation}
    \begin{aligned}
        &f_{12}(C)\quad\rightarrow\quad\frac{1}{\llangle23\rrangle\llangle34\rrangle^2}~,\qquad f_{34}(C)\quad\rightarrow\quad\frac{1}{\llangle23\rrangle\llangle34\rrangle^2}~,\\
        &f_{23}(C)\quad\rightarrow\quad\frac{\llangle12\rrangle}{\llangle23\rrangle^2\llangle34\rrangle^2}~,\qquad f_{14}(C)\quad\rightarrow\quad\frac{\llangle12\rrangle}{\llangle23\rrangle^2\llangle34\rrangle^2}~.
    \end{aligned}
\end{equation}
Since $C=\Lambda$, the remaining $\delta$-functions become
\begin{equation}
    \delta^6(C\Omega C^T)\delta^8(C\Omega \eta^T) \qquad \rightarrow \qquad \delta^6(\Lambda\Omega\Lambda^T)\delta^8(\Lambda\Omega\eta^T)=\delta^6(p)\delta^8(q)~.
\end{equation}
We can see that on the support of the momentum conserving $\delta$-functions all the non-trivial basic building blocks $I_{ij}$ multiplied with Mandelstams become
\begin{equation}
\label{result}
    \llangle ij\rrangle I_{ij}\quad \rightarrow \quad \frac{\delta(p)\delta(q)}{s_{12}s_{23}}~,
\end{equation}
which imply that $\gamma_{12}+\gamma_{23}+\gamma_{34}+\gamma_{14}=-i$ \cite{Plefka14, Huang,Dennen_Huang_Siegel10}. The fermionic $\delta$-function deserves a comment. According to the definition \eqref{ferm_delta_def} the fermionic $\delta$-function for 4-points now  reads
\begin{equation}
        \delta(q)=\prod_{I=1}^2\prod_{A=1}^4\delta\left(\sum_{i=1}^4\lambda^{Aa}_i\epsilon_{ab}\eta^{Ib}_i\right)~.
    \end{equation}
It can be shown with the help of  the definitions for dimensional reduction \eqref{dim_red_4D} and \eqref{eta_embeding} that individual supercharges dimensionally reduce to
 \begin{equation}
        q^{IA}_i|_{\textrm{4D~massive}}=
        \left(
        \begin{array}{cc}
             \eta^1_i\lambda_{i\alpha}-\tilde{\eta}_{i3}\mu_{i\alpha}&\eta^1_{i}\tilde{\mu}^{\dot{\alpha}}_i +\tilde{\eta}_{i3}\tilde{\lambda}_{i\dot{\alpha}} \\
             -\eta^4_{i}\lambda_{i\alpha}+\tilde{\eta}_{i2}\mu_{i\alpha}&-\eta^4_i\tilde{\mu}_i^{\dot{\alpha}}-\tilde{\eta}_{i2}\tilde{\lambda}^{\dot{\alpha}}_i
        \end{array}
        \right)~,
    \end{equation}
which agree with non-chiral supercharges for massive particles obtained in \cite{Plefka14}. Therefore, under dimensional reduction the 6D fermionic $\delta$-function correctly produces the product of both the 4D non-chiral $\delta$-functions $\delta^4(q)\delta^4(\tilde{q}) $, and the dimensionally reduced 4-point amplitude \eqref{result} agrees with results obtained from 6D $\mathcal{N}=(1,1)$ SYM \cite{Dennen_Huang_Siegel10}.

\section{Conclusions and future directions}

This paper has been dedicated to probe new ways to find a Grassmannian formula in terms of Pl\"{u}cker coordinates for scattering amplitudes of massive particles in 4D. In order to take advantage of massless kinematics, we worked with the chiral 6D model, which allowed us to use a symplectic Grassmannian that naturally talks to 6D kinematics. Although there are some issues coming from using a chiral model, 
it allows us to construct amplitudes with even number of external legs.

It turns out to be impossible to write massive amplitudes as a "pure" Grassmannian integral, because such formulas are naturally (super)conformally invariant while massive 4D amplitudes are not. The solution is to write the amplitude as a linear combination of 6D (super)conformally (potentially Yangian?) invariant Grassmannian integrals with momentum-dependent coefficients. It would be interesting to consider such expansion of amplitudes for other theories as well. The 6D dual conformal symmetry turned out to be a highly valuable tool. The modifications of dual conformal generators with all dual conformal weights equal (cf.\ massless 4D $\mathcal{N}=4$ SYM or 3D ABJM theory) has been generalized for the purposes of massive theories to the case of arbitrary dual conformal weights of scattering amplitudes. At this point emerged an interesting connection to the so-called evaluation representation of the Yangian algebra discussed by L.~Ferro et al. \cite{Ferro14}.

The (i) modified dilation of the dual conformal algebra and (ii) what we would call natural symmetries (momentum, Lorentz, little group, etc.) together with (iii) the internal $GL(n)$ symmetry of the symplectic Grassmannian strongly reduced the un-fixed degrees of freedom in the Grassmannian formula to a batch of theory-dependent numbers (entries of minors and powers of minors). Those can be further constrained with the help of the special dual conformal generator. This helped us to fix the 4-point Grassmannian formula up to (i) factors that on-shell gives 1, and (ii) proportionality constants, that cannot be determined from symmetry arguments.

The procedure discussed in this paper allows us to fix the 4-pt.\ Grassmannian formula and find candidates for theory-dependent integrands. The remaining information necessary for evaluation of the Grassmannian integrals is the integration contour. That is not necessary for 4-pt.\ example, because in that case the number of integrations equals the number of $\delta$-functions. In the context of $\mathcal{N}=4$ SYM, there are at least two possible approaches for finding the integration contour: i) introducing link variables \cite{Spradlin09} or ii) using on-shell diagrams to encode the BCFW \cite{Trnka12}. Both have been successfully used e.g.\ for 4D $\mathcal{N}=4$ SYM and $\mathcal{N}=8$ SUGRA \cite{Farrow-Lipstein-17}.  Similar methods could be in principle also be used in 6D, although it is challenging. In the first case we would have to know the world-sheet formulation of the pertinent theory. However, according to the  authors' knowledge, our model does not fit any known 6D model described by a world-sheet formulation \cite{Geyer18, Cachazo}. Furthermore, the usage of BCFW in our formulation could be tricky, because, as discussed above, we are not able to construct odd $n$-amplitudes relevant for 4D massive theory. A natural future direction is to adapt the approach discussed in this paper directly to 4D, which could solve the aforementioned issues and we could determine the contour, because in 4D there exist both the world-sheet formulation of massive $\mathcal{N}=4$ Coulomb branch amplitudes \cite{Schwarz} and all $n$-point amplitudes.

\appendix
\section{Appendix: Dirac $\delta$-distribution of antisymmetric matrix}
\label{Dirac-antisymm}

Given an antisymmetric $n\times n$ Grassmann-even matrix $A^I\equiv A^{i_1i_2}$, where $I\in\{1, \ldots, \frac{n(n-1)}{2}\}$ is an antisymmetric double-index. Define 1-parameter family as
\begin{equation}A(t)~:=~\Lambda(t)A\Lambda^T(t), \qquad \Lambda(t)~:=~e^{t\lambda}, \qquad t~\in~\mathbb{R}~. 
\end{equation} 

\vspace{5mm}
\noindent
{\em Lemma}:
\begin{equation}
\delta^{\frac{n(n-1)}{2}}(A(t))
~=~\frac{1}{|\det \Lambda(t)|^{(n-1)}}\delta^{\frac{n(n-1)}{2}}(A)~.
\end{equation}

\vspace{5mm}
\noindent
{\em Sketched proof of Lemma}: The derivative is
\begin{equation}
\frac{dA(t)}{dt} ~=~ \lambda A(t)+  A(t)\lambda^T
~=~ \Lambda(t)\left(\lambda A+  A\lambda^T\right)\Lambda^T(t), \qquad A(t\!=\!0)~=~A~. 
\end{equation}
The Dirac derivative \cite{6D_nase} is
\begin{equation}
\left(\frac{\partial A^{i_1i_2}}{\partial A^{j_1j_2}}\right)_{\! D}~=~\frac{1}{2}\left( \delta^{i_1}_{j_1}\delta^{i_2}_{j_2}
-(j_1\leftrightarrow j_2)\right)~.
\end{equation}
The Jacobian matrix is
\begin{equation}M^I{}_J(t)~:=~\left(\frac{\partial A^I(t)}{\partial A^J}\right)_{\! D}
~=~\frac{1}{2}\left( \Lambda^{i_1}{}_{j_1}(t) \Lambda^{i_2}{}_{j_2}(t)
-(j_1\leftrightarrow j_2)  \right)~,
\end{equation} 
\begin{equation}
\Lambda(t\!=\!0)~=~{\bf 1}_{n\times n},\qquad  M(t\!=\!0)~=~{\bf 1}_{\frac{n(n-1)}{2}\times \frac{n(n-1)}{2}}~.  
\end{equation}
The derivative is
\begin{equation} \frac{dM^I{}_J(t)}{dt}
~=~\frac{1}{2}\left(
\frac{d\Lambda^{i_1}{}_{j_1}(t)}{dt} \Lambda^{i_2}{}_{j_2}(t)
+\Lambda^{i_1}{}_{j_1}(t) \frac{d\Lambda^{i_2}{}_{j_2}(t)}{dt}
-(j_1\leftrightarrow j_2)  \right)~.
\end{equation}
The inverse Jacobian matrix is
\begin{equation}
(M^{-1})^J{}_I(t)~=~\frac{1}{2}\left( (\Lambda^{-1})^{j_1}{}_{i_1}(t) (\Lambda^{-1})^{j_2}{}_{i_2}(t)
-(i_1\leftrightarrow i_2)  \right)~. 
\end{equation}
We compute:
\begin{equation}
\begin{aligned} \ln \det M(t\!=\!1)
~=~&\int_0^1 \! dt ~\frac{d}{dt}\ln \det M(t) \cr
~=~&\int_0^1 \! dt ~\frac{d}{dt}{\rm tr} \ln M(t) \cr
~=~&\int_0^1 \! dt ~{\rm tr}\left( M^{-1}(t) \frac{dM(t)}{dt} \right)\cr
~=~&\int_0^1 \! dt ~ (M^{-1})^J{}_I(t) \frac{dM^I{}_J(t)}{dt} \cr
~=~&\frac{1}{4}\int_0^1 \! dt ~\left( (\Lambda^{-1})^{j_1}{}_{i_1}(t) (\Lambda^{-1})^{j_2}{}_{i_2}(t)
-(i_1\leftrightarrow i_2)  \right)  \cr
&\times \left(
\frac{d\Lambda^{i_1}{}_{j_1}(t)}{dt} \Lambda^{i_2}{}_{j_2}(t)
+\Lambda^{i_1}{}_{j_1}(t) \frac{d\Lambda^{i_2}{}_{j_2}(t)}{dt}
-(j_1\leftrightarrow j_2)  \right)\cr
~=~&\ldots \cr
~=~&(n\!-\!1)\int_0^1 \! dt ~ (\Lambda^{-1})^j{}_i(t) \frac{d\Lambda^i{}_j(t)}{dt} \cr
~=~&(n\!-\!1)\int_0^1 \! dt ~{\rm tr}\left( \Lambda^{-1}(t) \frac{d\Lambda(t)}{dt} \right)\cr
~=~&(n\!-\!1)\int_0^1 \! dt ~\frac{d}{dt}{\rm tr} \ln \Lambda(t) \cr
~=~&(n\!-\!1)\int_0^1 \! dt ~\frac{d}{dt}\ln \det \Lambda(t) \cr
~=~&(n\!-\!1)\ln \det \Lambda(t\!=\!1)~.
\end{aligned}
\end{equation}
$\Box$

\section{Appendix: The action of $K^{AD}$ on $\langle l^a~l_a~m^b~m_b \rangle$}
\label{app_K_on_angle}
We here give a detailed calculation of the action of $K^{AD}$ on the Lorentz invariant angle-bracket $\langle l^a~l_a~m^b~m_b\rangle$. We  begin with the last expression in eq.\ \eqref{K-1}
\begin{equation}
\label{neco}
   K^{AD}\langle l^a~l_a~m^b~m_b\rangle= -\frac{1}{4}\left(\sum_{j<i}-\sum_{j>i}\right)\lambda^{[Ac}_{j}\lambda^{D]d}_i\Lambda^{\mathcal{B}}_{jc}\partial_{id\mathcal{B}}~\langle l^a~l_a~m^b~m_b\rangle~.
\end{equation}
First of all, it's enough to consider just the bosonic part of the operator, because angle brackets depend on $\lambda$ only, 
\begin{equation}
    \lambda^{[Ac}_{j}\lambda^{D]d}_i\lambda^{B}_{jc}\partial_{idB}~\langle l^a~l_a~m^b~m_b\rangle=2\left(\delta_{il}p^{[AE}_jp_l^{D]F}p_{m}^{GH}\epsilon_{EFGH}+ \delta_{im}p_j^{[AG}p_m^{D]H}p_l^{EF}\epsilon_{EFGH}\right)~.
\end{equation}
Eq.\ \eqref{neco} therefore continues as
\begin{equation}
\label{pomoc_1}
\begin{aligned}
    =&-\frac{1}{2}\left(\sum_{j<i}-\sum_{j>i}\right)\left[\delta_{il}p_j^{[AE}p_l^{D]F}p_m^{GH}\epsilon_{EFGH}+\delta_{im}p_j^{[AE}p_m^{D]F}p_l^{GH}\epsilon_{EFGH}\right]
\\
    =&-\frac{1}{2}\left(\sum_{j<l}-\sum_{j>l}\right)p_j^{[AE}p_l^{D]F}p_m^{GH}\epsilon_{EFGH}-\frac{1}{2}\left(\sum_{j<m}-\sum_{j>m}\right)p_j^{[AE}p_m^{D]F}p_m^{GH}\epsilon_{EFGH}~.
\end{aligned}
\end{equation}
The above expression can be written in terms of angle brackets with help of the formula
\begin{equation}
    p_j^{[AE}p_l^{D]F}p_m^{GH}\epsilon_{EFGH}=\frac{1}{2}\left[p_{j}^{AD}\langle l^a~l_a~m^b~m_b\rangle-p_m^{AD}\langle j^a~j_a~l^b~l_b\rangle+p_l^{AD}\langle j^a~j_a~m^b~m_b\rangle\right]~,
\end{equation}
which follows from repeated use of the Schouten identity \cite{Plefka14}
\begin{equation}
    p^{[AE}_jp^{D]F}_lp^{GH}_m\epsilon_{EFGH}=p^{AD}_j\langle l^a~l_a~m^b~m_b\rangle-p_j^{[AE}p^{D]H}_mp^{MF}_l\epsilon_{EHMF}~.
\end{equation}
The expression \eqref{pomoc_1} now becomes
\begin{equation}
\label{pomoc2}
\begin{aligned}
=-\frac{1}{4}\left(\sum_{j<l}-\sum_{j>l}\right)&\left[p_j^{AD}\langle l^a~l_a~m^b~m_b\rangle-p_m^{AD}\langle j^a~j_a~l^b~l_b\rangle+p_l^{AD}\langle j^a~j_a~m^b~m_b\rangle\right]\\
-\frac{1}{4}\left(\sum_{j<m}-\sum_{j>m}\right)&\left[p_j^{AD}\langle l^a~l_a~m^b~m_b\rangle-p^{AD}_l\langle j^a~j_a~m^b~m_b\rangle+p^{AD}_m\langle j^a~j_a~l^b~l_b\rangle\right]~.
\end{aligned}
\end{equation}
We can now rewrite the terms in \eqref{pomoc2} that contain the same angle brackets. E.g.\ the sum of the first term in each of the 2 square brackets becomes
    \begin{equation*}
    \begin{aligned}
        -&\frac{1}{4}\left(\sum_{j<l}-\sum_{m\geq j>l} -\sum_{j>m}+\sum_{j<l}+\sum_{m>j\geq l}-\sum_{j>m}\right)p_j^{AD}\langle l^a~l_a~m^b~m_b\rangle\\
        =&-\frac{1}{2}\left(\sum_{j<l}-\sum_{j>m}\right)p_j^{AD}\langle l^a~l_a~m^b~m_b\rangle+\frac{1}{4}p_m^{AD}\langle l^a~l_a~m^b~m_b\rangle-\frac{1}{4}p^{AD}_l\langle l^a~l_a~m^b~m_b~. \rangle
        \end{aligned}
    \end{equation*}
The other terms pair up similarly. The final result for \eqref{pomoc2} is
\begin{equation}
=-\frac{1}{2}\left(\sum_{j<l}-\sum_{j>m}\right)p_j^{AD}\langle l^a~l_a~m^b~m_b\rangle
-\frac{1}{2}\sum_{m>j>l}\left[p_m^{AD}\langle j^a~j_a~l^b~l_b\rangle-p_l^{AD}\langle j^a~j_a~m^b~m_b\rangle\right]~.
\end{equation}

\section{Appendix: The action of $K^{AD}$ on $(ij\ldots l)$}
\label{K_on_minor}

\subsection{Non-gauge fixed version}
\label{non-gauge-fixed}
Let's now investigate the action of the operator \eqref{K_useful} on the 4-point minors. We begin with the first term in \eqref{K_useful} only
\begin{equation}
\label{bla1}
    -\frac{1}{4}\sum_{j<i}\Lambda^{[Ab}_j\Lambda^{D]a}_i\mathcal{O}_{jb|ia}(xy)=\frac{1}{4}\sum_{j<i}\Lambda^{[Ab}_j\Lambda^{D]a}_i\sum_{p=1}^n\epsilon_{ac}C^{pc}_i\frac{\partial}{\partial C^{pb}_j}(xy)~.
\end{equation}
The action of the second term goes analogously. Applying the replacement operator \eqref{lambdacreplacement} on the minor yields
\begin{equation}
\label{K_on_minor_I}
    \sum_{p=1}^nC^{p}_{ia}\frac{\partial}{\partial C^{pb}_j}(xy)=\sum_{p=1}^nC^{p}_{ia}\frac{\partial}{\partial C^{pb}_j}\left[\epsilon_{rstu}C^{re}_x\epsilon_{ef}C^{sf}_xC^{tg}_y\epsilon_{gh}C^{uh}_y\right]~=
\end{equation}
\begin{equation*}
    =\delta_{xj}(i_ax_by^gy_g)-\delta_{xj}(x_bi_ay^gy_g)+\delta_{yj}(x^ex_ei_ay_b)-\delta_{yj}(x^ex_ey_bi_a)~.
\end{equation*}
Although the first two terms are identical, we prefer to keep this form to make the Schouten identities obvious.

We split now the 4-point calculation into two cases: \textit{consecutive} 4-pt.\ case and \textit{non-consecutive} 4-pt.\ case. Firstly, we consider \textbf{consecutive minors} where $|x-y|=1$. Here we have three sub-cases:
\begin{itemize}
    \item $i\leq\textrm{min}(x,y)$, then the derivative in \eqref{K_on_minor_I} does not hit the minor and therefore it is 0.
    
    \item $\textrm{min}(x,y)< i\leq\textrm{max}(x,y)$, e.g. $i=y$
    \begin{equation}
    \label{three_cases}
        \ldots \left[(i_ax_bi^gi_g)-(x_bi_ai^gi_g)\right]~=~0~.
    \end{equation}
    This vanishes, because there will be always two identical columns.
    
    \item $i>\textrm{max}(x,y)$ is the only part that contributes.
\end{itemize}

\noindent
We calculate
\begin{equation}
    =\frac{1}{4}\sum_{i=\textrm{max}(x,y)+1}^n\Lambda^{[Da}_i\left[\Lambda^{A]b}_x(i_ax_by^gy_g)-\Lambda^{A]b}_x(x_bi_ay^gy_g)+\Lambda^{A]b}_y(x^ex_ei_ay_b)-\Lambda^{A]b}_y(x^ex_ey_bi_a)\right]~.
\end{equation}
We can now use the following completeness relation
\begin{equation}
\label{compleetness_rel}
    \delta_{ij}\delta^d_a=\sum_{m=1}^n\left[(\hat{C}^T)_{ia}^{m}C^{md}_{j}-(C^T)^m_{ia}\hat{C}^{md}_j\right]~,
\end{equation}
which will be proven in the subsection \ref{proof_compl_rel}, and express all $\Lambda_x$ and $\Lambda_y$ as
\begin{equation}
\label{lambda_in_C}
    \Lambda^{Ab}_x=\sum_{l=1}^n\Lambda^{Ae}_l\delta_{xl}\delta^b_e=\sum_{l,m=1}^n\Lambda^{Ae}_l(\hat{C}^T)^m_{le}C^{mb}_x~,
\end{equation}
where the second term from eq.\ \eqref{compleetness_rel} vanishes on the support of the  $\delta(C\Omega\Lambda^T)$. Thus we get
\begin{equation}
\begin{aligned}
    =\frac{1}{4}\sum_{i=\textrm{max}(x,y)+1}^n\Lambda^{[Da}_i\sum_{l=1}^n\sum_{m=1}^n\Lambda^{A]e}_l(\hat{C}^T)^m_{le}\left[C^{mb}_x(i_ax_by^gy_g)-C^{mb}_x(x_bi_ay^gy_g)\right.\\
    \left.+C^{mb}_y(x^ex_ei_ay_b)-C^{mb}_y(x^ex_ey_bi_a)\right]
\end{aligned}
\end{equation}
\begin{equation}
    =\ldots\left[-C^{m}_{xb}(i_ay^gy_gx^b)-C^{mb}_x(x_bi_ay^gy_g)-C^{m}_{yb}(x^ex_ei_ay^b)-C^{mb}_y(y_bx^ex_ei_a)\right]~.
\end{equation} We now use an $n$-term Schouten identity and get
\begin{equation}
    =\frac{1}{4}\sum_{i=\textrm{max}(x,y)+1}^n\Lambda^{[Da}_i\sum_{l=1}^n\sum_{m=1}^n\Lambda^{A]e}_l(\hat{C}^T)^m_{le}C^m_{ia}(x^ex_ey^gy_g)~.
\end{equation}
Next use the completeness relation \eqref{compleetness_rel}
\begin{equation}
    =\frac{1}{4}\sum_{i=\textrm{max}(x,y)+1}^n\Lambda^{[Da}_i\Lambda^{A]}_{ia}(x^ex_ey^gy_g)=-\frac{1}{2}\sum_{i>\textrm{max}(x,y)}p^{AD}_i(xy)~.
\end{equation}
Therefore we can conclude that the first term gives
\begin{equation}
    -\frac{1}{4}\sum_{j<i}\Lambda^{[Ab}_j\Lambda^{D]a}_i\mathcal{O}_{jb|ia}(xy)=-\frac{1}{2}\sum_{i>\textrm{max}(x,y)}p^{AD}_i (xy)~,
\end{equation}
and similarly for the second term in $K$
\begin{equation}
    -\frac{1}{4}\sum_{j<i}\Lambda^{[Ab}_j\Lambda^{D]a}_i\mathcal{O}_{ia|jb}(xy)=\frac{1}{2}\sum_{j<\textrm{min}(x,y)}p^{AD}_j (xy)~.
\end{equation}
To conclude, the action of the $K^{AD}$ on consecutive minors can be written as
\begin{equation}
    K^{AD}(xy)=-\frac{1}{2}\left(\sum_{i>\textrm{max}(x,y)}-\sum_{i<\textrm{min}(x,y)}\right)p_i^{AD}(xy)~.
\end{equation}

Secondly, let us consider \textbf{non-consecutive minors}, where $|x-y|\geq2$. We begin by reviewing the three different cases in \eqref{three_cases}.

\begin{itemize}
    \item If $i\leq\textrm{min}(x,y)$, then the derivative in \eqref{K_on_minor_I} does not hit the minor and therefore it is 0.
    
    \item If $\textrm{min}(x,y)< i\leq\textrm{max}(x,y)$, then we have a new contribution
    \begin{equation}
        \frac{1}{4}\sum_{i}\Lambda^{[Da}_i\left[\Lambda^{A]b}_x(i_a~x_b~y^g~y_g)-\Lambda^{A]b}_x(x_b~i_a~y^g~y_g)\right]~.
    \end{equation}

    \item If $i>\textrm{max}(x,y)$, we  have the same contribution.
\end{itemize}
Finally, the boundary term vanishes
\begin{equation}
    i=y:\qquad (y_a~x_b~y^g~y_g)=0~.
\end{equation}
Therefore the only new contribution from the $\mathcal{O}_{jb|ia}$ part is
\begin{equation}
    \frac{1}{4}\sum_{x<i<y}\Lambda^{[Da}_i\left[\Lambda^{A]b}_x(i_a~x_b~y^g~y_g)-\Lambda^{A]b}_x(x_b~i_a~y^g~y_g)\right]~,
\end{equation}
and from $\mathcal{O}_{ia|jb}$
\begin{equation}
    \frac{1}{4}\sum_{x<j<y}\Lambda^{[Ab}_j\left[\Lambda^{D]a}_y(x^e~x_e~j_b~y_a)-\Lambda^{D]a}_y(x^e~x_e~y_a~j_b)\right]~.
\end{equation}
After a little manipulation we get a complete formula 
\begin{equation}
\label{K_on_minor_nooneighbours}
     \frac{1}{4}\sum_{x<i<y}\Lambda^{[Da}_i\left[\Lambda^{A]b}_x(i_a~x_b~y^g~y_g)-\Lambda^{A]b}_x(x_b~i_a~y^g~y_g)-\Lambda^{A]b}_y(x^e~x_e~i_a~y_b)+\Lambda^{A]b}_y(x^e~x_e~y_b~i_a)\right]~.
\end{equation}
Next step is to use the formula \eqref{compleetness_rel} and the Schouten identities. Omitting sums and prefactors for clarity, the first two terms in \eqref{K_on_minor_nooneighbours} yield
\begin{equation}
    -\Lambda^{[Ab}_x\left[\Lambda^{D]}_{xb}(y^g~y_g~i^a~i_a)+\Lambda^{D]g}_y(y_g~i^a~i_a~x_b)+\Lambda^{D]}_{yg}(i^a~i_a~x_b~y^g)\right]~,
\end{equation}
and the last two terms in \eqref{K_on_minor_nooneighbours} are
\begin{equation}
    \Lambda^{[Ab}_y\left[\Lambda^{D]e}_x(x_e~y_b~i^a~i_a)+\Lambda^{D]}_{xe}(y_b~i^a~i_a~x^e)+\Lambda^{D]}_{yb}(i^a~i_a~x^e~x_e)\right]~.~~~
\end{equation}
The new contribution for 4-pt.\ non-consecutive minors is
\begin{equation}
    =\frac{1}{2}\sum_{x<i<y}\left[p^{AD}_y(i~x)-p_x^{AD}(i~y)\right]~.
\end{equation}
All together, the action of the special dual conformal generator on a general 4-pt.\ minor is
\begin{equation}
\label{end}
    K^{AD}(xy)=-\frac{1}{2}\left(\sum_{i>y}-\sum_{i<x}\right)p_i^{AD}(xy)-\frac{1}{2}\sum_{x<i<y}\left[p_x^{AD}(iy)-p_y^{AD}(ix)\right]~.
\end{equation}

\subsection{Gauge fixed version}

We give here a gauge fixed version of the proof in appendix \ref{non-gauge-fixed} for the case $n=4$ investigated in this paper. This will lead to a new contributions in \eqref{end}. The considered gauge fixed $C$ matrix is of the form where first $n$ columns are gauge fixed to an orthonormal basis
\begin{equation}
\label{gauge_fixing}
    C^{m1}_k=\delta^m_k~\ldots\textrm{ gauge~fixed}, \qquad C^{m2}_k  ~\ldots\textrm{ variable}.
\end{equation}

Let us begin with the generator \eqref{K_useful}
\begin{equation}
     K^{AD}=-\frac{1}{4}\sum_{j<i}\Lambda^{[Ab}_j\Lambda^{D]a}_i\left(\epsilon_{bc}\Lambda^{\mathcal{A}c}_{j}\frac{\partial}{\partial \Lambda^{\mathcal{A}a}_{i}}+\epsilon_{ac}\Lambda^{\mathcal{A}c}_{i}\frac{\partial}{\partial \Lambda^{\mathcal{A}b}_{j}}\right)~.
\end{equation}
This generator acts on the product of $\delta$-functions, which in the gauge fixed case \eqref{gauge_fixing} becomes 
\begin{equation}
    \prod_{m=1}^n\prod_{\mathcal{W}=1}^{4|\mathcal{N}}\delta\left(\sum_{k=1}^n C^{ma}_k\epsilon_{ab}\Lambda^{\mathcal{W}b}_{k}\right)~=~\prod_{m=1}^n\prod_{\mathcal{W}=1}^{4|\mathcal{N}}\delta\left(\Lambda^{\mathcal{W}2}_m-\sum_{k=1}^n C^{m2}_k\Lambda^{\mathcal{W}1}_k\right)~.
\end{equation}
We now repeat the idea, where operators of the form $\Lambda\frac{\partial}{\partial \Lambda}$ are replaced by some first-order differential operator acting on $C$. Due to the combination of little group indices, it splits into four cases
\begin{equation}
\label{replacement_rules}
    \begin{aligned}
        \Lambda^{\mathcal{A}1}_j\frac{\partial}{\partial \Lambda^{\mathcal{A}1}_i}  \qquad \rightarrow&\qquad \sum_{p=1}^n C^{p2}_i\frac{\partial}{\partial C^{p2}_j}~,\\
        \Lambda^{\mathcal{A}1}_j\frac{\partial}{\partial \Lambda^{\mathcal{A}2}_i}  \qquad \rightarrow&\qquad -\frac{\partial}{\partial C^{i2}_j}~=~-\sum_{p=1}^nC^{p1}_i\frac{\partial}{\partial C^{p2}_j}~,\\
        \Lambda^{\mathcal{A}2}_j\frac{\partial}{\partial \Lambda^{\mathcal{A}1}_i}  \qquad\rightarrow&\qquad (-1)^{|\mathcal{A}|}\delta^{\mathcal{A}}_{\mathcal{A}} C_i^{j2}+\sum_{p,q=1}^nC^{j2}_qC^{p2}_i\frac{\partial}{\partial C^{p2}_q}~,\\
        \Lambda^{\mathcal{A}2}_j\frac{\partial}{\partial \Lambda^{\mathcal{A}2}_i}  \qquad\rightarrow&\qquad -\sum_{p=1}^nC^{j2}_p\frac{\partial}{\partial C^{i2}_p}~.
    \end{aligned}
\end{equation}
We can see that there appears a new term in the third line of \eqref{replacement_rules}, which will produce new contributions  as compared to the 3D and 4D versions of the proof \cite{Bargheer14, Drummond-Ferro_10}.  Thus the complete gauge fixed version of $K^{AD}$ acting on $C$ is
\begin{equation}
\label{K_in_C}
    \begin{aligned}
        K^{AD}~=~&-\frac{\epsilon_{12}}{4}\sum_{j<i}\Lambda^{[A1}_j\Lambda_i^{D]1}\sum_{q}\left[C^{j2}_q\sum_{p}C^{p2}_i\frac{\partial}{\partial C^{p2}_q}+C^{i2}_q\sum_pC^{p2}_j\frac{\partial}{\partial C^{p2}_q}\right]\\
        &+\frac{\epsilon_{12}}{4}\sum_{j<i}\Lambda^{[A1}_j\Lambda_i^{D]2}\left[\sum_{p}C^{j2}_p\frac{\partial}{\partial C^{i2}_p}+\sum_pC^{p2}_j\frac{\partial}{\partial C^{p2}_i}\right]\\
         &+\frac{\epsilon_{12}}{4}\sum_{j<i}\Lambda^{[A2}_j\Lambda_i^{D]1}\left[\sum_{p}C^{p2}_i\frac{\partial}{\partial C^{p2}_j}+\sum_pC^{i2}_p\frac{\partial}{\partial C^{j2}_p}\right]\\
         &-\frac{\epsilon_{12}}{4}\sum_{j<i}\Lambda^{[A2}_j\Lambda_i^{D]2}\left[\sum_{p}C^{p1}_i\frac{\partial}{\partial C^{p2}_j}+\sum_{p}C^{p1}_j\frac{\partial}{\partial C^{p2}_i}\right]~.
    \end{aligned}
\end{equation}
The operator \eqref{K_in_C} now acts on $\delta(C\Omega \Lambda^T)$ only and we need to integrate by parts to hit the function $f^{6D}(C)$. Then the operator hits the $\delta$-function $\delta(C\Omega C^T)$ also. It is easy to see that the last three lines in \eqref{K_in_C} annihilate the gauge fixed $\delta(C\Omega C^T)$, however, the first line deserves a deeper discussion. The action of the $C$-part of the first line in \eqref{K_in_C} on the gauge fixed $\delta$-function $\delta(C\Omega C^T)$ reads
\begin{equation}
\label{K_on_CC}
\begin{aligned}
     &\sum_{p,q}\left(C^{j2}_qC^{p2}_i+C^{i2}_qC^{p2}_j\right)\frac{\partial}{\partial C^{p2}_q}\prod_{l<m}\delta(C^{l2}_m-C^{m2}_l)~=\\
     =\sum_{p<q}[C^{j2}_qC^{p2}_i+&C^{i2}_qC^{p2}_j-C^{j2}_pC^{q2}_i-C^{i2}_pC^{q2}_j]\delta^{\prime}(C^{p2}_q-C^{q2}_p)\prod_{\substack{l<m \\ (l,m)\neq (p,q)}}\delta(C^{l2}_m-C^{m2}_l)~.
\end{aligned}
\end{equation}
There are now three sub-cases depending on whether the indices $p,q$ that we sum over coincide with the indices $i,j$ or not:
\begin{itemize}
    \item $p,q \notin \{i,j\}$\\
     The square bracket in \eqref{K_on_CC} is 0 on the support of undifferentiated $\delta$-functions.
     
    \item $p ~\textrm{or}~q\in\{i,j\}$, e.g. $p=j$
    \begin{equation}
    \begin{aligned}
        &C^{j2}_i(C^{j2}_q-C^{q2}_j)\delta^{\prime}(C^{j2}_q-C^{q2}_j)\prod_{\substack{l<m \\ (l,m)\neq (j,q)}}\delta(C^{l2}_m-C^{m2}_l)\\
        &=-C^{i2}_i\prod_{l<m}\delta(C^{l2}_m-C^{m2}_l)~.
    \end{aligned}
    \end{equation}
    
    \item $(p,q)=(i,j)$
    \begin{equation}
    \label{third}
    \begin{aligned}
        &(C^{j2}_i+C^{i2}_j)(C^{j2}_i-C^{i2}_j)\delta^{\prime}(C^{j2}_i-C^{i2}_j)\prod_{\substack{l<m \\ (l,m)\neq (j,i)}}\delta(C^{l2}_m-C^{m2}_l)\\
        &= -2C^{j2}_i\prod_{l<m}\delta(C^{l2}_m-C^{m2}_l)~.
    \end{aligned}
    \end{equation}
\end{itemize}
We can see that there will be one contribution of the type \eqref{third}, so the problem reduces to the question "How many times do $p$ or $q$ coincide with $i$ or $j$?" It can be shown that the answer is $2n-4$. Altogether the three sub-cases give $(2n-2)$ contributions and we have
\begin{equation}
    \frac{\epsilon_{12}}{4}\sum_{j<i}\Lambda^{[A1}_j\Lambda_i^{D]1}(2n-2)C^{j2}_i\delta(C\Omega C^T)~.
\end{equation}

Next we consider the case when $K^{AD}$ hits the function $f^{6D}(C)$. We assume the ansatz \eqref{f_ansatz}, i.e.\ the function $f^{6D}(C)$ depends on minors of $C$ and has a rational form. Thus we have to investigate the action of $K^{AD}$ on minors. This is however now much simpler, because not all of them are independent. It can be shown that for gauge fixed $C$, it holds that 
\begin{equation}
    (12)=(34)~, \quad (23)=(14)~, \quad (13)=(24)~, \quad (12)+(13)+(14)=0~.
\end{equation}
Therefore there are only two independent minors. We choose $(12)$ and $(23)$ to be independent. It is sufficient to investigate the action of $K^{AD}$ on these two minors only. Let us now for convenience introduce a shorthand notation for minors\footnote{
Note that the previous definition \eqref{minor_def} is $-4$ times the new definition \eqref{new_minor_def}.} 
\begin{equation}
\label{new_minor_def}
    (xy)\sim |C^1_xC^1_yC^2_xC^2_y|~=~\epsilon_{stuv}C^{s1}_xC^{t1}_yC^{u2}_xC^{v2}_y,
\end{equation}
where |....| denotes a determinant and $C^a_i$ is the $i$th column in the gauge fixed part of $C$ if $a=1$ and in the non-gauge fixed part of $C$ if $a=2$.

We can now investigate the action of \eqref{K_in_C} on the minor $(12)$. We can see that \eqref{K_in_C} contains combinations of $\Lambda$s with both little group indices. Our strategy will be to remove all little group indices 2 from the $\Lambda$s with the help of relation \eqref{lambda_in_C} and the Schouten identity for determinants.  Thus the action of the last three lines of \eqref{K_in_C} on $(12)$ can be simplified with the help of \eqref{lambda_in_C} and the Schouten identity. It takes a form
\begin{equation}
\label{last_three_lanes}
    -\frac{1}{2}\sum_{i>2}p_i^{AD}|C^1_1C^1_2C^2_1C^2_2|-\frac{\epsilon_{12}}{4}\sum_{i>2}\Lambda^{[D1}_i\left[\Lambda^{A]1}_1|C^1_2C^2_iC^2_2C^2_1|+\Lambda^{A]1}_2|C^2_iC^2_2C^2_1C^1_1|\right]~,
\end{equation}
where we recognize the first term, which is exactly what we would expect from \eqref{end}, while the last two terms will contribute to the action of the first line of \eqref{K_in_C} because of the little group structure on $\Lambda$s. We should further point out that minors in the bracket are basically third-order polynomials in the remaining variables of $C$. 

The action of the first line of \eqref{K_in_C} on (12) can be written as 
\begin{equation}
    \begin{aligned}
        -\frac{\epsilon_{12}}{4}\sum_{j<i}\Lambda^{[A1}_j\Lambda^{D]1}_i\left[C^{j2}_1|C^1_1C^1_2C^2_iC^2_2|\rule[-6pt]{0pt}{18pt}\right.+&~C^{j2}_2|C^1_1C^1_2C^2_1C^2_i|\\
        +C^{i2}_1|C^1_1C^1_2C^2_jC^2_2| +&\left. \rule[-6pt]{0pt}{18pt}C^{i2}_2|C^1_1C^1_2C^2_1C^2_j|\right]~,
    \end{aligned}
\end{equation}
where we should note that although minors inside the square bracket are second-order polynomials, due to $C$s in front of them, they are actually third-order polynomials. An Laplace row/column expansion of the determinant can be used on \eqref{last_three_lanes},  

\begin{equation}
    \begin{aligned}
        |C^1_2C^2_iC^2_2C^2_1|&=\left|
        \begin{array}{cccc}
             0&C^{12}_i&C^{12}_2&C^{12}_1  \\
             1&C^{22}_i&C^{22}_2&C^{22}_1  \\
             0&C^{32}_i&C^{32}_2&C^{32}_1  \\
             0&C^{42}_i&C^{42}_2&C^{42}_1
        \end{array}
        \right|~=~(-1)^{1+2}C^{12}_i\left|\begin{array}{ccc}
             1&C^{22}_2&C^{22}_1  \\
             0&C^{32}_2&C^{32}_1  \\
             0&C^{42}_2&C^{42}_1
        \end{array}
        \right|+\ldots\\
        &=C^{12}_i\left|
        \begin{array}{cccc}
             0&1&C^{12}_2&C^{12}_1  \\
             1&0&C^{22}_2&C^{22}_1  \\
             0&0&C^{32}_2&C^{32}_1  \\
             0&0&C^{42}_2&C^{42}_1
        \end{array}
        \right|+\ldots
        \\&=C^{12}_i~|C^1_2C^1_1C^2_2C^2_1|+C^{12}_2|C^1_2C^2_iC^1_1C^2_1|+C^{12}_1|C^1_2C^2_iC^2_2C^1_1|~,
    \end{aligned}
\end{equation}
and similarly
\begin{equation}
        |C^2_iC^2_2C^2_1C^1_1|=C^{22}_i|C^1_2C^2_2C^2_1C^1_1|+C^{22}_2|C^2_iC^1_2C^2_1C^1_1|+C^{22}_1|C^2_iC^2_2C^1_2C^1_1|~,
\end{equation}
which establishes a connection between these two contributions. An explicit calculation shows that the result of the little group $(1,1)$ term is
\begin{equation}
    -\frac{\epsilon_{12}}{4}\sum_{j<i}\Lambda^{[A1}_j\Lambda^{D]1}_i(C^{j2}_i+C^{i2}_j)(12)~,
\end{equation}
which is a new contribution. The complete action of $K^{AD}$ on the gauge fixed minor (12) has the form
\begin{equation}
\label{vysledek_1}
    K^{AD}(12)=-\frac{1}{2}\sum_{i>2}p_i^{AD}(12)-\frac{\epsilon_{12}}{4}\sum_{j<i}\Lambda^{[A1}_j\Lambda^{D]1}_i(C^{j2}_i+C^{i2}_j)(12)~.
\end{equation}
Similar calculation holds also for the (23) minor
\begin{equation}
\label{vysledek_2}
    K^{AD}(23)=-\left(\frac{1}{2}p^{AD}_4-\frac{1}{2}p^{AD}_1\right)(23)-\frac{\epsilon_{12}}{4}\sum_{j<i}\Lambda^{[A1}_j\Lambda^{D]1}_i(C^{j2}_i+C^{i2}_j)(23)~.
\end{equation}

We can see that the gauge fixed calculations \eqref{vysledek_1} and \eqref{vysledek_2} differ from the previous non-gauge fixed calculation \eqref{end} by a term proportional to  $\Lambda^{[A1}_j\Lambda^{D]1}_i$. There is, however, one more term of the form $\frac{\partial}{\partial C}CC$ that needs to be taken into account, as will become clear from the following calculation. Let us consider the following integral, where we assume no boundary terms
\begin{equation}
    \begin{aligned}
            0=&-\frac{\epsilon_{12}}{4}\sum_{j<i}\Lambda^{[A1}_j\Lambda^{D]1}_i\int\textrm{d}C\sum_{p,q}\frac{\partial}{\partial C^{p2}_q}\left[\left(C^{j2}_qC^{p2}_i+C^{i2}_qC^{p2}_j\right)f^{6D}_{lm}(C)\delta(C\Omega C^T)\delta(C\Omega\Lambda^T)\right]=\\
        =&-\frac{\epsilon_{12}}{4}\sum_{j<i}\Lambda^{[A1}_j\Lambda^{D]1}_i\int\textrm{d}C\left[\sum_{p,q}\frac{\partial}{\partial C^{p2}_q}\left(C^{j2}_qC^{p2}_i+C^{i2}_qC^{p2}_j\right)\right]f^{6D}_{lm}(C)\delta(C\Omega C^T)\delta(C\Omega\Lambda^T)+\\
        &+\int\textrm{d}C\left[K^{AD}_{11}f^{6D}_{lm}(C)\right]\delta(C\Omega C^T)\delta(C\Omega\Lambda^T)+\int\textrm{d}Cf^{6D}_{lm}(C)\left[K^{AD}_{11}\delta(C\Omega C^T)\right]\delta(C\Omega\Lambda^T)\\
        &+\int\textrm{d}Cf^{6D}_{lm}(C)\delta(C\Omega C^T)\left[K^{AD}_{11}\delta(C\Omega\Lambda^T)\right]~.
    \end{aligned}
\end{equation}
Plugging all the previous results leads to 
\begin{equation}
    \begin{aligned}
        &-\frac{\epsilon_{12}}{4}\sum_{j<i}\Lambda^{[A1}_j\Lambda^{D]1}_i\int\textrm{d}C\left[4n-(2n-2)+2(a_{lm}+\ldots+f_{lm})\right]C^{j2}_if^{6D}(C)\delta(C\Omega C^T)\delta(C\Omega\Lambda^T)\\
        &+\int\textrm{d}Cf^{6D}(C)\delta(C\Omega C^T)\left[K^{AD}_{11}\delta(C\Omega\Lambda^T)\right]~=~0~.
    \end{aligned}
\end{equation}
Combining aforementioned equation with the contact term in the third line of eq.\ \eqref{replacement_rules} we get the final  $\Lambda^{[A1}_j\Lambda^{D]1}_i$ extra contribution 
\begin{equation}
\label{vysledek_3}
    -\frac{\epsilon_{12}}{4}\sum_{j<i}\Lambda^{[A1}_j\Lambda^{D]1}_i\int\textrm{d}C\left[(3-\mathcal{N}-n)-(a_{lm}+\ldots+f_{lm})\right]C^{j2}_if^{6D}(C)\delta(C\Omega C^T)\delta(C\Omega\Lambda^T)~,
\end{equation}
which vanishes, because the $GL(n)$ scaling \eqref{homogenity} tells us that  $a_{lm}+\ldots+f_{lm}=3-\mathcal{N}-n$ (or equivalently, it is implied by dual dilaton invariance). To summarize, all the extra contributions in the gauge fixed calculation \eqref{vysledek_3} vanish and the equations implied by $K^{AD}$ are the same compared to the non-gauge fixed case \eqref{12sector} and \eqref{23sector}.

\subsection{Proof of completeness relation \eqref{compleetness_rel}}
\label{proof_compl_rel}

We give a proof of the completeness relation \eqref{compleetness_rel} and  will be using the double index notation, where Greek indices ($\mu$, $\nu$, etc.) runs from 1 to $2n$ and represents tuple $(i,a)$. Although we are interested in 6D, the proof is similar to the 3D case \cite{Lee10}. We assume that we have a given symplectic Grassmannian $C$, i.e.
\begin{equation}
\label{geometric_condition}
     C^{m}{}_{\mu}\Omega^{\mu\nu} (C^T)_{\nu}{}^{n}=0~,
\end{equation}
where $C$ can be viewed as an $n\times 2n$ matrix. Let us now define an auxiliary symplectic Grassmannian  $\hat{C}^{m}{}_{\mu}$ that is related to original Grassmannian $C^m{}_{\mu}$ by invertible matrix $A^{mn}$ 
\begin{equation}
\label{defining_equations}
    \begin{aligned}
    \hat{C}^{m}{}_{\nu}\Omega^{\nu\mu}(\hat{C}^T)_{\mu}{}^{n}&=0~, \qquad
    C^m{}_{\mu}\Omega^{\mu\nu}(\hat{C}^T)_{\nu}{}^{n}=A^{mn}~.
    \end{aligned}
\end{equation}
Strictly speaking to prove \eqref{compleetness_rel} we need just $A^{mn}=\mathbb{1}^{mn}$. However we can further generalize this simple case to arbitrary symmetric invertible matrix $A^{mn}$, whose number of d.o.f.\ matches with the dimension of the symplectic Grassmannian. This by itself does not guarantee existence of the auxiliary Grassmannian $\hat{C}$. Let us prove the existence of $\hat{C}$ for complementary gauge fixing of Grassmannian defines as follows
\begin{equation}
\label{complementary_gaugefix}
\begin{aligned}
   &~~~~~~~~~~~~~~~C_{n\times 2n}=(E_{n\times n}F_{n\times n})\\ 
    &~E~~~\left\{\begin{aligned}
        ~~~\textrm{Orthonormal gauge fixed}:&\qquad \{i_1, \ldots , i_k\}\\
   \textrm{ Non-gauge fixed}:&\qquad \{j_1,\ldots, j_{n-k}\}
    \end{aligned}\right.\\ 
    &~F~~~\left\{
    \begin{aligned}
        ~~~\textrm{Orthonormal gauge fixed}:&\qquad \{j_1, \ldots , j_{n-k}\}\\
   \textrm{ Non-gauge fixed}:&\qquad \{i_1,\ldots, i_{k}\}
    \end{aligned}
    \right.
\end{aligned}
\end{equation}
The gauge fixing \eqref{complementary_gaugefix} linearizes the quadratic symplectic constraint \eqref{geometric_condition}. This can be easily seen if we write the constraint \eqref{geometric_condition} in terms of $E$ and $F$. The symplectic constraint becomes
\begin{equation}
    EF^T=FE^T~,
\end{equation}
and we see that the gauge fixed part of $E$ always hits non-gauge fixed part of matrix $F$ and vice-versa. Thus the condition is now linear. Let us now gauge fix the auxiliary Grassmannian $\hat{C}$ in the same way as $C$. Then eqs.\ \eqref{defining_equations} become linear and we can always find $\hat{C}$ for complementary gauge fixing \eqref{complementary_gaugefix}.

With the help of the matrix $A^{mn}$ we can define projection operators $P^{\mu\nu}$ and $\bar{P}^{\mu\nu}$
\begin{equation}
    P_{\rho\mu}:=(\hat{C}^T)_{\rho}{}^p(A^{-1})_{pm}C^m{}_{\mu}~, \qquad \bar{P}_{\rho\sigma}:=(C^T)_{\rho}{}^{s}(A^{-1})_{sm}\hat{C}^m{}_{\sigma}~,
\end{equation}
satisfying
\begin{equation}
    \begin{aligned}
    P_{\rho\mu}\Omega^{\mu\nu}P_{\nu\sigma}=P_{\rho\sigma}~, \qquad & \bar{P}_{\rho,\mu}\Omega^{\mu\nu}\bar{P}_{\nu\sigma}=\bar{P}_{\rho\sigma}~,\\
    \\
    P_{\rho\mu}\Omega^{\mu\nu}\bar{P}_{\nu\sigma}=\bar{P}_{\rho\mu}&\Omega^{\mu\nu}P_{\nu\sigma}=0~.
    \end{aligned}
\end{equation}
It remains to show that 
the difference of projection operators is equal to $\Omega_{\mu\nu}$. This can be seen by defining the sum
\begin{equation}
    D_{\mu\nu}:=P_{\mu\nu}-\bar{P}_{\mu\nu}~,
\end{equation}
and multiplying with $C$ and $\hat{C}$ 
\begin{equation}
\begin{aligned}
    C^{n}{}_{\rho}\Omega^{\rho\mu}D_{\mu\nu}=C^{n}{}_{\nu}~, \qquad D_{\mu\nu}\Omega^{\nu\rho}(C^T)_{\rho}{}_{l}=(C^T)_{\mu}{}^{l}~,\\
\hat{C}^n{}_{\rho}\Omega^{\rho\mu}D_{\mu\nu}=\hat{C}^n{}_{\nu}~,\qquad D_{\mu\nu}\Omega_{\nu\rho}(\hat{C}^T)_{\rho}{}^l=(\hat{C}^T)_{\mu}{}_{l}~,
\end{aligned}
\end{equation}
which imply
\begin{equation}
    D_{\mu\nu}=\Omega_{\mu\nu}~.
\end{equation}

\section{Appendix: $n=4$ calculation details}
\label{n=4_details} 
We give an intermediate step of the $n=4$ calculation in this appendix. The required independent equations \eqref{12sector} and \eqref{23sector} were derived in  section \ref{example}.  However, it is more insightful to use the linear combination \eqref{12sector}-\eqref{23sector} instead of \eqref{23sector}, which reads
\begin{equation}
    \begin{aligned}
    -&\frac{1}{2}[\gamma_{12}f_{12}(C)+\gamma_{34}f_{34}(C)-\gamma_{13}f_{13}(C)-\gamma_{24}f_{24}(C)](3p_1^{AD}+2p_2^{AD}-p_4^{AD})\\
    +&\frac{1}{2}[\gamma_{23}f_{23}(C)+\gamma_{14}f_{14}(C)-\gamma_{13}f_{13}(C)-\gamma_{24}f_{24}(C)](4p_1^{AD}+2p_2^{AD}+p_3^{AD}-p_4^{AD})\\
    -&\gamma_{12}\left[K^{AD}f_{12}(C)\right]-\gamma_{34}\left[K^{AD}f_{34}(C)\right]+\gamma_{23}\left[K^{AD}f_{23}(C)\right]+\gamma_{14}\left[K^{AD}f_{14}(C)\right]\\
    =&(p_1^{AD}+p_2^{AD}+p_3^{AD}+p_4^{AD})\left[l_{12}\gamma_{12}f_{12}(C)+l_{34}\gamma_{34}f_{34}(C)-l_{23}\gamma_{23}f_{23}(C)-l_{14}\gamma_{14}f_{14}(C)\right]~.
    \end{aligned}
\end{equation}
This equation implies 4 equations according to $p_i^{AD}$:
\begin{itemize}
    \item $p_1^{AD}$
    \begin{equation}
    \begin{aligned}
    &-\frac{3}{2}(\gamma_{12}f_{12}(C)+\gamma_{34}f_{34}(C))+2(\gamma_{23}f_{23}(C)+\gamma_{14}f_{14}(C))-\frac{1}{2}(\gamma_{13}f_{13}(C)+\gamma_{24}f_{24}(C))\\
    &-\gamma_{12}\frac{f_{12}(C)}{2}\left[b_{12}+c_{12}+f_{12}-d_{12}\frac{(23)}{(13)}-e_{12}\frac{(24)+(34)}{(14)}\right]\\ 
    &-\gamma_{34}\frac{f_{34}(C)}{2}\left[b_{34}+c_{34}+f_{34}-d_{34}\frac{(23)}{(13)}-e_{34}\frac{(24)+(34)}{(14)}\right] \\
    &+\gamma_{23}\frac{f_{23}(C)}{2}\left[b_{23}+c_{23}+f_{23}-d_{23}\frac{(23)}{(13)}-e_{23}\frac{(24)+(34)}{(14)}\right]\\ 
    &+\gamma_{14}\frac{f_{14}(C)}{2}\left[b_{14}+c_{14}+f_{14}-d_{14}\frac{(23)}{(13)}-e_{14}\frac{(24)+(34)}{(14)}\right]\\
    &=l_{12}\gamma_{12}f_{12}(C)+l_{34}\gamma_{34}f_{34}(C)-l_{23}\gamma_{23}f_{23}(C)-l_{14}\gamma_{14}f_{14}(C)~.
    \end{aligned}
    \end{equation}
    
    \item $p^{AD}_2$
    \begin{equation}
        \begin{aligned}
        &-(\gamma_{12}f_{12}(C)+\gamma_{34}f_{34}(C))+(\gamma_{23}f_{23}(C)+\gamma_{14}f_{14}(C))\\
        &-\gamma_{12}\frac{f_{12}(C)}{2}\left[c_{12}-f_{12}\frac{(34)}{(24)}\right]-\gamma_{34}\frac{f_{34}(C)}{2}\left[c_{34}-f_{34}\frac{(34)}{(24)}\right]\\
        &+\gamma_{23}\frac{f_{23}(C)}{2}\left[c_{23}-f_{23}\frac{(34)}{(24)}\right]+\gamma_{14}\frac{f_{14}(C)}{2}\left[c_{14}-f_{14}\frac{(34)}{(24)}\right]\\
        &=l_{12}\gamma_{12}f_{12}(C)+l_{34}\gamma_{34}f_{34}(C)-l_{23}\gamma_{23}f_{23}(C)-l_{14}\gamma_{14}f_{14}(C)~.
        \end{aligned}
    \end{equation}
    
    \item $p^{AD}_3$
    \begin{equation}
        \begin{aligned}
            &\frac{1}{2}\left(\gamma_{23}f_{23}(C)+\gamma_{14}f_{14}(C)-\gamma_{13}f_{13}(C)-\gamma_{24}f_{24}(C)\right)\\
            &-\gamma_{12}\frac{f_{12}(C)}{2}\left[-a_{12}+d_{12}\frac{(12)}{(13)}\right]-\gamma_{34}\frac{f_{34}(C)}{2}\left[-a_{34}+d_{34}\frac{(12)}{(13)}\right]\\
            &+\gamma_{23}\frac{f_{23}(C)}{2}\left[-a_{23}+d_{23}\frac{(12)}{(13)}\right]+\gamma_{14}\frac{f_{14}(C)}{2}\left[-a_{14}+d_{14}\frac{(12)}{(13)}\right]\\
            &=l_{12}\gamma_{12}f_{12}(C)+l_{34}\gamma_{34}f_{34}(C)-l_{23}\gamma_{23}f_{23}(C)-l_{14}\gamma_{14}f_{14}(C)~.
        \end{aligned}
    \end{equation}
    
    \item $p_4^{AD}$
    \begin{equation}
        \begin{aligned}
        &\frac{1}{2}(\gamma_{12}f_{12}(C)+\gamma_{34}f_{34}(C)-\gamma_{23}f_{23}(C)-\gamma_{14}f_{14}(C))\\
        &-\gamma_{12}\frac{f_{12}(C)}{2}\left[-a_{12}-b_{12}-d_{12}+f_{12}\frac{(23)}{(24)}+e_{12}\frac{(12)+(13)}{(14)}\right]\\
        &-\gamma_{34}\frac{f_{34}(C)}{2}\left[-a_{34}-b_{34}-d_{34}+f_{34}\frac{(23)}{(24)}+e_{34}\frac{(12)+(13)}{(14)}\right]\\
        &+\gamma_{23}\frac{f_{23}(C)}{2}\left[-a_{23}-b_{23}-d_{23}+f_{23}\frac{(23)}{(24)}+e_{23}\frac{(12)+(13)}{(14)}\right]\\
        &+\gamma_{14}\frac{f_{14}(C)}{2}\left[-a_{14}-b_{14}-d_{14}+f_{14}\frac{(23)}{(24)}+e_{14}\frac{(12)+(13)}{(14)}\right]\\
        &=l_{12}\gamma_{12}f_{12}(C)+l_{34}\gamma_{34}f_{34}(C)-l_{23}\gamma_{23}f_{23}(C)-l_{14}\gamma_{14}f_{14}(C)~.
        \end{aligned}
    \end{equation}
    
\end{itemize}

\noindent Similarly for the equation \eqref{12sector}
\begin{itemize}
    \item $p_1^{AD}$
    \begin{equation}
        \begin{aligned}
            &-\frac{3}{2}\left(\gamma_{12}f_{12}(C)+\gamma_{34}f_{34}-\gamma_{13}f_{13}(C)-\gamma_{24}f_{24}(C)\right)\\
            &-\gamma_{12}\frac{f_{12}(C)}{2}\left[b_{12}+c_{12}+f_{12}-d_{12}\frac{(23)}{(13)}-e_{12}\frac{(24)+(34)}{(14)}\right]\\
            &-\gamma_{34}\frac{f_{34}(C)}{2}\left[b_{34}+c_{34}+f_{34}-d_{34}\frac{(23)}{(13)}-e_{34}\frac{(24)+(34)}{(14)}\right]\\
            &+\gamma_{13}\frac{f_{13}(C)}{2}\left[b_{13}+c_{13}+f_{13}-d_{13}\frac{(23)}{(13)}-e_{13}\frac{(24)+(34)}{(14)}\right]\\
            &+\gamma_{24}\frac{f_{24}(C)}{2}\left[b_{24}+c_{24}+f_{24}-d_{24}\frac{(23)}{(13)}-e_{24}\frac{(24)+(34)}{(14)}\right]\\
            &=l_{12}\gamma_{12}f_{12}(C)+l_{34}\gamma_{34}f_{34}(C)-l_{13}\gamma_{13}f_{13}(C)-l_{24}\gamma_{24}f_{24}(C)~.
        \end{aligned}
    \end{equation}
    
    \item $p_2^{AD}$
    \begin{equation}
        \begin{aligned}
            &-(\gamma_{12}f_{12}(C)+\gamma_{34}f_{34}(C)-\gamma_{13}f_{13}(C)-\gamma_{24}f_{24}(C))\\
            &-\gamma_{12}\frac{f_{12}(C)}{2}\left[c_{12}-f_{12}\frac{(34)}{(24)}\right]-\gamma_{34}\frac{f_{34}(C)}{2}\left[c_{34}-f_{34}\frac{(34)}{(24)}\right]\\
            &+\gamma_{13}\frac{f_{13}(C)}{2}\left[c_{13}-f_{13}\frac{(34)}{(24)}\right]+\gamma_{24}\frac{f_{24}(C)}{2}\left[c_{24}-f_{24}\frac{(34)}{(24)}\right]\\
            &=l_{12}\gamma_{12}f_{12}(C)+l_{34}\gamma_{34}f_{34}(C)-l_{13}\gamma_{13}f_{13}(C)-l_{24}\gamma_{24}f_{24}(C)~.
        \end{aligned}
    \end{equation}
    
    \item $p_3^{AD}$
    \begin{equation}
        \begin{aligned}
            &-\gamma_{12}\frac{f_{12}(C)}{2}\left[-a_{12}+d_{12}\frac{(12)}{(13)}\right]-\gamma_{34}\frac{f_{34}(C)}{2}\left[-a_{34}+d_{34}\frac{(12)}{(13)}\right]\\
            &+\gamma_{13}\frac{f_{13}(C)}{2}\left[-a_{13}+d_{13}\frac{(12)}{(13)}\right]+\gamma_{24}\frac{f_{24}(C)}{2}\left[-a_{24}+d_{24}\frac{(12)}{(13)}\right]\\
            &=l_{12}\gamma_{12}f_{12}(C)+l_{34}\gamma_{34}f_{34}(C)-l_{13}\gamma_{13}f_{13}(C)-l_{24}\gamma_{24}f_{24}(C)~.
        \end{aligned}
    \end{equation}
    
    \item $p^{AD}_4$
    \begin{equation}
        \begin{aligned}
            &\frac{1}{2}(\gamma_{12}f_{12}(C)+\gamma_{34}f_{34}(C)-\gamma_{13}f_{13}(C)-\gamma_{24}f_{24}(C))\\
            &-\gamma_{12}\frac{f_{12}(C)}{2}\left[-a_{12}-b_{12}-d_{12}+f_{12}\frac{(23)}{(24)}+e_{12}\frac{(12)+(13)}{(14)}\right]\\
             &-\gamma_{34}\frac{f_{34}(C)}{2}\left[-a_{34}-b_{34}-d_{34}+f_{34}\frac{(23)}{(24)}+e_{34}\frac{(12)+(13)}{(14)}\right]\\
              &+\gamma_{13}\frac{f_{13}(C)}{2}\left[-a_{13}-b_{13}-d_{13}+f_{13}\frac{(23)}{(24)}+e_{13}\frac{(12)+(13)}{(14)}\right]\\
               &+\gamma_{24}\frac{f_{24}(C)}{2}\left[-a_{24}-b_{24}-d_{24}+f_{24}\frac{(23)}{(24)}+e_{24}\frac{(12)+(13)}{(14)}\right]\\
            &=l_{12}\gamma_{12}f_{12}(C)+l_{34}\gamma_{34}f_{34}(C)-l_{13}\gamma_{13}f_{13}(C)-l_{24}\gamma_{24}f_{24}(C)~.
        \end{aligned}
    \end{equation}
\end{itemize}

\acknowledgments

The authors thank Yu-tin Huang, Aidan Herderschee and Arthur Lipstein for helpful discussions, Patrik Novosad for feedback on a draft of the manuscript and  the anonymous referee for his/her insightful and helpful comments. The work of K.B. is supported by the Czech Science Foundation (GACR) under the grant no. GA20-04800S for Integrable Deformations. Computational resources were supplied by the MetaCentrum project "e-Infrastruktura CZ" (e-INFRA CZ LM2018140) supported by the Ministry of Education, Youth and Sports of the Czech Republic.

\end{document}